\documentclass[aps,amsmath,preprint,epsf,superscriptaddress,nofootinbib]{revtex4-1}
\usepackage{graphicx}
\usepackage{bm}
\usepackage{color}
\usepackage{amsmath,amssymb}	
\usepackage{hyperref}
\usepackage{setspace}
\usepackage{longtable}
\usepackage{booktabs}

\pdfoutput=1

\begin{document}
%%%%%%%%%%%%%%%%%%%%%%%%%%%%%%%%%%
%%%%%%%%%%% Title page %%%%%%%%%%%
%%%%%%%%%%%%%%%%%%%%%%%%%%%%%%%%%%

\preprint{KEK-TH-1913}
\preprint{IPMU16-0092}
\bigskip

\title{{\Large Cracking Down on Fake Photons }\\ \sl{-- A Case of 750\,GeV Diphoton Resonance --}}

\author{Hajime Fukuda}
\email[e-mail: ]{hajime.fukuda@ipmu.jp}
\affiliation{Kavli IPMU (WPI), UTIAS, The University of Tokyo, Kashiwa, Chiba 277-8583, Japan}

\author{Masahiro Ibe}
\email[e-mail: ]{ibe@icrr.u-tokyo.ac.jp}
\affiliation{Kavli IPMU (WPI), UTIAS, The University of Tokyo, Kashiwa, Chiba 277-8583, Japan}
\affiliation{ICRR, The University of Tokyo, Kashiwa, Chiba 277-8582, Japan}

\author{Osamu Jinnouchi}
\email[e-mail: ]{jinnouchi@phys.titech.ac.jp}
\affiliation{Tokyo Institute of Technology, Tokyo, 152-8551, Japan}

\author{Mihoko Nojiri}
\email[e-mail: ]{nojiri@post.kek.jp}
\affiliation{KEK Theory Center, IPNS, KEK, Tsukuba, 305-0801, Japan}
\affiliation{Graduate University of Advanced Studies (Sokendai),Tsukuba, 305-0801, Japan}
\affiliation{Kavli IPMU (WPI), UTIAS, The University of Tokyo, Kashiwa, Chiba 277-8583, Japan}

\date{\today}

\begin{abstract}
Among various models to explain the $750$\,GeV diphoton resonance hinted at the LHC Run 2,
a class of models where the resonance decays not into a pair of photons but into a pair of 
photon-jets is gathering definite attention.
In this paper, we study how well we can distinguish the di-photon-jet resonance 
from the diphoton resonance by examining detector responses 
to the photon-jets.  We find that the  sum of $p_T$ of the  first $e^+e^-$ pair from the photon conversion
provides strong discrimination power.
We also discuss determination of the lifetime of the light intermediate particle by measuring the photon conversion points. 
\end{abstract}

\maketitle

%%%%%%%%%%%%%%%%%%%%%%%%%%%%%%%%%%%%
%%%%%%%%%%% Introduction %%%%%%%%%%%
%%%%%%%%%%%%%%%%%%%%%%%%%%%%%%%%%%%%
\section{Introduction}
The ATLAS and the CMS collaborations reported an intriguing hint for a new resonance in 
diphoton invariant mass spectrum around $750$\,GeV~\cite{ATLAS:2015aa,CMS:2015dxe, collaboration:2016aa,CMS:2016owr}.
Although more integrated luminosity is required to determine whether
the resonance is from a real signal, a plethora of models have been proposed to account for the resonance after the reports~\cite{diphoton:2016}.

For obvious reasons, most of models involve a  neutral scalar boson 
with a mass around $750$\,GeV which decays into a pair of photons.
(See e.g. Refs.~\cite{Knapen:2015dap,Buttazzo:2015txu,Franceschini:2015kwy,McDermott:2015sck,Ellis:2015oso,Low:2015qep,Gupta:2015zzs,Dutta:2015wqh,
Agrawal:2015dbf,Berthier:2015vbb,Falkowski:2015swt,Alves:2015jgx,Kim:2015ksf,Craig:2015lra} for early phenomenological interpretations.)
Since the required production cross section to explain the signal is rather large,
the resonance is assumed to be  produced by the gluon fusion in most of the models.
There, the gluon fusion process and the decay of the resonance into a pair of photons 
are induced by loop diagrams of new colored/charged particles.
 
 As alternative possibilities, models with the resonance decaying not into a pair of photons but into a pair of 
photon-jets also gather definite 
attention~\cite{Knapen:2015dap,Agrawal:2015dbf,Higaki:2015jag,Chang:2015sdy,Chala:2015cev,Aparicio:2016iwr,Ellwanger:2016qax,Domingo:2016unq,Chiang:2016eav,Alves:2016koo}.
When the photons constituting the photon-jets are highly collimated, they are difficult to be distinguished from isolated 
single photons by the electromagnetic calorimeters (ECAL). 
In this way, the di-photon-jet resonance can fake the signature of the diphoton resonance.

A phenomenological advantage of this class of the models is that 
the resonance can decay into the photon-jets via tree-level diagrams. 
Thus, the decay width and the branching ratio into a pair of the photon-jets can be much larger
than those in the models where the resonance decays into a pair of photons via loop diagrams.%
\footnote{This feature might be a further advantage of this class of models 
in view of the tentative hints for the large decay width of the resonance.
}
Accordingly,  the signal cross section can be explained with a fewer number of the new colored/charged particles 
compared with the models with the  diphoton signature.

From theoretical point of view, this class of the models is also motivated to provide  appropriate masses 
to the new colored/charged particles.
For example, let us assume that the masses of the new colored particles in the TeV range are induced 
by spontaneous symmetry breaking of an approximate chiral symmetry.
In this case, the models are naturally associated with a pseudo Nambu-Goldstone boson as well as a scalar particle which consist of 
a complex scalar field whose vacuum expectation value (VEV) is responsible for the chiral symmetry breaking.
In this class of models, the $750$\,GeV resonance can be identified with the radial component of the complex field.
Then, it dominantly decays into a pair of 
pseudo Nambu-Goldstone bosons which subsequently decay into photons.
Therefore, all the ingredients necessary for the di-photon-jet interpretation of the $750$\,GeV resonance
are incorporated.

In this paper, we study how well we can distinguish  di-photon-jet resonances from a diphoton resonance 
by the development ofthe electromagnetic shower of the photon candidates by using simplified detector simulations.
So far, it has been discussed that the different conversion rates of the photons and the photon-jets into $e^+e^-$ pairs
provide strong discriminating variables to reject the photon-jets interpretation~\cite{Ellwanger:2016qax,Dasgupta:2016wxw}.
In our study, we further investigate the possibility to measure the $p_T$ distribution of the 
converted photons which are reconstructed by measuring the $e^+e^-$ track momenta. The converted di-photon-jet signature is associated with a
lower  $p_T$ electron pair than converted diphoton signature. The errors of the momentum of electron tracks from the photon conversion are not equal to that of electron tracks from interaction points as one cannot use full volume of trackers. We estimate the distribution after the smearing by making a reasonable assumption to the position dependent momentum resolution of the electron pair. In addition, we perform simplified simulation of electromagnetic shower evolution in the ATLAS and CMS inner trackers to take into account electron bremsstrahlung effects, which is essential to estimate the contamination of diphoton signature 
into di-photon-jet signal regions.   
We also discuss whether we can extract the finite decay length of the light particle by analyzing the photon-conversion points.

The organization of the paper is as follows.
In section\,\ref{sec:model}, we briefly review the models in which the $750$\,GeV resonance decaying into photon-jets.
In section\,\ref{sec:detector}, we summarize the behavior of the photon-jets inside the detectors.
In section\,\ref{sec:photonjets}, we show our analysis on how well we can distinguish the photon-jets from isolated photons.
We find the di-photon-jet signature can be distinguished from 
diphoton signature at  integrated luminosity ${\cal L}_0= 25\,\text{fb}^{-1}$ if the signal cross section after the selection cut is above 
2\,fb.  In section\,\ref{sec:lifetime}, we discuss whether we can study the lifetime of the intermediate particle.
Final section is devoted to our conclusions.  

\section{The $750$\,GeV resonance decaying into photon-jets}
\label{sec:model}
In this section, we briefly summarize the models where the $750$\,GeV resonance is decaying into photon-jets.
In particular, we take a simple example where this class of models naturally emerges 
as a result of spontaneous breaking of an approximate chiral symmetry. 

%%%%%%%%%%%%%%%%%%%%%%%%%%%%%%%%%%%%%%%%%%%%%%%%%%%%%%%%%%%%%%%%%%
\subsection{Signal cross section, branching ratios of the resonance}
First, let us briefly review models of the diphoton resonance where the scalar resonance is produced via the gluon fusion process
and decays into a pair of photons through a loop diagrams.
This class of models can be achieved by, for example, introducing $N_f$-flavors of new colored fermions ($\psi$,$\bar{\psi}$).
The new fermions are assumed to couple to the scalar resonance $s$ with a mass $M_s \simeq 750$\,GeV via
\begin{eqnarray}
	{\cal L} = M_F \bar{\psi}\psi + \frac{g}{\sqrt{2}}s \bar{\psi} \psi + h.c. \,
\end{eqnarray}
where $M_F$ and $g$ denote a mass parameter and a coupling constant, respectively. 
Here, we assume that all the $\psi$'s are in the fundamental representations of QCD and have the QED charge $Q$
for simplicity.

In this setup, the resonance is produced by the gluon fusion at the LHC through 
loop diagrams in which the new colored particles are circulating.
The production cross section of this process is roughly given by,
\begin{eqnarray}
	\label{eq:production}
	\sigma(pp\to s) \simeq 6\,{\rm fb} \times 
	N_f^2\left(\frac{1\,\rm TeV}{M_F/g}\right)^2
	\left(\frac{f(t_F)}{2/3}\right)^2\ ,
\end{eqnarray}
where we use the {\tt MSTW2008} parton distribution functions~\cite{Martin:2009iq}.%
\footnote{Here, we do not include the so called $K$-factor of ${\cal O}(1)$ describing higher order corrections, 
	which is not very relevant for  following discussions.}
The function $f(t)$ (with $t_F= 4M_F^2/M_s^2$) is defined by,
\begin{eqnarray}
	\label{eq:f}
	f(t)\equiv t \left[ 1 + (1-t) \arcsin^2(1/\sqrt{t}) 
	\right]\ ,
\end{eqnarray}
which immediately converges to $f(t)\to 2/3$ for $t \gg 1$.

By compared with the global fits of the signal cross sections~\cite{Franceschini:2016gxv}, 
\begin{eqnarray}
	\label{eq:required}
	\sigma(pp\to s)\times Br(s\to 2 \gamma) &=& (5.5 \pm 1.5)\, {\rm fb} \quad {\rm (ATLAS)} \ ,  \\
	\sigma(pp\to s)\times Br(s\to 2 \gamma) &=& (4.8 \pm 2.1)\, {\rm fb} \quad  {\rm (CMS)} \ ,
\end{eqnarray}
the signal requires $Br(s \to \gamma\gamma) = {\cal O}(1)$ for $M_F \simeq 1$\,TeV, $g\simeq 1$ and $N_f = \mathcal O(1)$.
As the resonance $s$ decays into a pair of photons via loop diagrams, however, 
$Br(s \to 2\gamma)$ is typically suppressed by ${\cal O}(\alpha_{\rm QED}^2/\alpha_{s}^2)$. 
Here, $\alpha_{\rm QED}$ and $\alpha_{s}$ are the fine-structure constants of  QED and  QCD, respectively.
Therefore, in this class of the models, the signal requires either $M_F \ll 1$\,TeV or large $N_f $ or $Q$~\cite{Aguilar-Saavedra:2013qpa,Ellis:2014dza} 
(see also Ref.\,\cite{Bae:2016xni} for constraints from perturbativity on the coupling constants).

Now, let us move on to the di-photon-jet scenario.
In this scenario, $s$  decays not into a pair of photons but mainly into a pair of a light scalar particles $a$.
The light scalar particle subsequently decays into multiple ($N_\gamma$) photons. 
When the mass of $a$, $m_a$, is smaller than about $1$\,GeV, the multiple photons from the decay of $a$ 
are highly collimated and form a narrow photon-jet, which is identified as a photon signal by the ECAL.
In this case, the signal cross section is provided by
\begin{eqnarray}
	\sigma(pp\to s)\times Br(s\to 2 a ) \times Br(a\to N_\gamma \gamma)^2  \ .
\end{eqnarray}
Thus, the required cross section in Eq.\,(\ref{eq:required}) can be easily obtained for
$Br(s\to 2 a ) = {\cal O}(1)$ and $Br(a\to N_\gamma \gamma) = {\cal O}(1)$ 
even for $M_F \simeq 1$\,TeV,  $N_f = 1$ and $Q = 1$, for example,
which is advantageous from the view point of model building.
In the following, we discuss a simple model which realizes such branching ratios.

%%%%%%%%%%%%%%%%%%%%%%%%%%%%%%%%%%%
\subsection{A model based on chiral symmetry breaking}
As in the model in the previous section, let us introduce a pair of colored left-handed Weyl fermions ($\psi$, $\bar{\psi}$) 
in the fundamental and anti-fundamental representations of $SU(3)_c$.
This time, we assume that the masses of the vector-like fermions are suppressed by an approximate chiral $U(1)$ symmetry.
Instead, the new fermions couple to a complex scalar field $\phi$ which has an appropriate chiral $U(1)$ charge via
\begin{eqnarray}
	\label{eq:Yukawa}
	{\cal L}_{\rm int} = g \phi \bar{\psi}\psi + h.c.\  ,
\end{eqnarray}
with $g$ being a coupling constant.

Now, let us assume that the chiral symmetry is broken spontaneously by the VEV of $\phi$ at around the TeV scale.
Then, the fermions obtain a mass $M_F = g\langle{\phi}\rangle$ in the TeV range for $g \simeq 1$.
The advantage of this type of the models is that it has all the necessary ingredients for the di-photon-jet scenario.
In fact, 
we may decompose $\phi$ into a scalar boson $s$ and a pseudo Nambu-Goldstone boson $a$
at around the VEV of $\phi$, $\langle{\phi}\rangle =f_a/\sqrt{2}$; %
\footnote{In terms of the pseudo Nambu-Goldstone mode, the chiral symmetry is non-linearly realized by $a/f_a \to a/f_a +  \alpha$  ($\alpha \in [0,2\pi)$).}
\begin{eqnarray}
	\phi = \frac{1}{\sqrt{2}}(f_a + s) e^{i a/f_a}\ .
	\label{eq:decompose}
\end{eqnarray}
The mass of $s$ is expected to be around ${\cal O}(f_a)$ while 
the mass of $a$ is much smaller, i.e. $m_a \ll M_s$, due to the approximate chiral symmetry. 
Therefore, $s$ can be reasonably identified with the $750$\,GeV resonance for $f_a  = {\cal O}(100)$\,GeV--${\cal O}(1)$\,TeV while, 
$a$ can play a role of the intermediate particle decaying into a photon-jet.
In the following, we call the pseudo Nambu-Goldstone, the axion-like particle (ALP).%
\footnote{In the model discussed in \cite{Chiang:2016eav}, the ALP can be the QCD axion which solves the strong $CP$ problem.}

Below the mass scale of $M_F$, the effective interactions involving $s$ and $a$ are given by
\begin{eqnarray}
	\label{eq:eff1}
	{\mathcal L}_{\rm eff} &=& \frac{s}{f_a}\partial_\mu a\partial^\mu a 
	+ \frac{\alpha_sN_f}{8\pi}\frac{g}{\sqrt{2}M_F}f\left(t_F\right) 
	sGG
	+\frac{\alpha_sN_f}{8\pi}\frac{a}{f_a} G\tilde{G}\ ,
\end{eqnarray}
where $G$ denotes the field strength of QCD (with suppressed Lorentz indices)  and $f(t_F)$ is given in Eq.\,(\ref{eq:f}).%
\footnote{In our discussion, we are assuming that the couplings of $\phi$ to the Higgs doublets in the Standard 
	Model are rather suppressed.} 
Here, we assume that the vector-like fermions do not carry other gauge charges under 
the $SU(2)_L\times U(1)_Y$ gauge groups.
When the vector-like fermions are $SU(2)_L\times U(1)_Y$ charged, $s$ and $a$ also couple
to the $SU(2)_L\times U(1)_Y$  gauge bosons, which does not affect the following arguments 
significantly.

From the effective interaction terms in Eq.\,(\ref{eq:eff1}), we  immediately find that the scalar resonance $s$
mainly decays into a pair of the ALPs with a decay rate
\begin{eqnarray}
	\label{eq:dilaton_width}
	\Gamma(s\to 2 a) \simeq \frac{{M_s}^3}{32\pi{f_a}^2} \simeq 4.2\,\text{GeV}\times
	\left(
	\frac{1\,\rm TeV}{f_a}
	\right)^2
\end{eqnarray}
for $M_s = 750\,\text{GeV}$. 
The coupling of $s$ with the gluons is, on the other hand, suppressed by a loop factor, and hence, 
it provides subdominant decay width of $s$ into jets, while it provides the  gluon fusion production cross section as in Eq.\,(\ref{eq:production}).
As a result, the model satisfies one of the requirement, $Br(s\to 2 a) \simeq 1$.

Decay properties of the ALP require  more careful studies.
Here, we follow the discussion given in Ref.~\cite{Chiang:2016eav}.
At a first glance, the main decay modes of the ALP seem to be the ones into jets or hadrons 
since the ALP only couples to the gluons in the effective Lagrangian.
In order for the photon-jets to be highly collimated, however, the mass of the ALP is required to 
be at most about $1$\,GeV.
For such a light state, the phase space of the ALP decay into the hadrons is highly limited, 
and hence, the modes into the hadrons are suppressed.
In fact, as shown in Ref.~\cite{Chiang:2016eav}, 
the ALP mainly decays to light mesons and photons through the mixing to the $\eta$ and the $\eta'$ mesons 
in the Standard Model.

To discuss the ALP mixings with the $\eta$ and the $\eta'$ mesons, let us consider the effective Lagrangian 
below the chiral symmetry breaking scale of QCD,
\begin{eqnarray}
	{\cal L}_{\rm mix} &=& -\frac{1}{2}m_a^2\, a^2 
	- \frac{1}{2}m_8^2\, \eta_8^2 - {\mit \Delta}^2 \eta_8\eta_0
	- \frac{1}{2}m_0^2\left(\eta_0 + N_f\frac{f_0}{\sqrt{6}f_a} a\right)^2 \ ,
	\label{eq:eff0}
\end{eqnarray} 
where $\eta_{8,0}$ denote the neutral component of the octet 
and the singlet pseudo-NG modes in QCD, respectively.
The parameter $f_0$ denotes the decay constant of $\eta_0$.
In the mass range we are interested, the mixing to $\pi^0$ is found to be negligible.

The mass parameters $m_8^2$ and ${\mit \Delta}^2$ represent the explicit 
breaking of the chiral symmetry of QCD by the quark mass terms 
%\begin{eqnarray}
%m_8^2 &\simeq & \frac{1}{3}\frac{(m_u + m_d + 4m_s)}{(m_u+m_d)} m_\pi^2\ , \\
%{  \mit\Delta}^2 &\simeq & \frac{\sqrt{2}}{3}\frac{(m_u + m_d -2m_s)}{(m_u+m_d)} m_\pi^2\ ,
%\label{eq:CHPT}
%\end{eqnarray}
(see, {\it e.g.}, Ref.~\cite{Beisert:2003zs}).
%The mass parameters of the quarks and pions should be understood.
The ALP mass $m_a$ encapsulates  explicit breaking of the approximate chiral $U(1)$ symmetry 
of the vector-like fermion other than the QCD anomaly.
Finally, the parameter $m_0^2$ represents the explicit breaking 
of both the $U(1)_A$ symmetry of QCD and the chiral symmetry of the vector-like colored fermions by QCD anomalies.%
\footnote{The $\eta_0$ meson is normalized such that the $U(1)_A$ symmetry is realized 
	\begin{eqnarray}
		\eta_0/f_0 \to \eta_0/f_0 + \alpha\,\quad &\longleftrightarrow& \quad q_{L,R} \to (e^{i\alpha/\sqrt{6}},\, e^{i\alpha/\sqrt{6}},\,e^{i\alpha/\sqrt{6}})\times q_{L,R}  \ ,
	\end{eqnarray}
	where $q_{L,R} = (u_{L,R}, d_{L,R}, s_{L,R})$ are the quarks in the Standard Model.
}
The ALP is mixed with $\eta_{0,8}$ through this term.

Once the ALP is mixed with $\eta_{0,8}$, it decays into a pair of photons though
the anomalous coupling of $\eta_{0,8}$ to the photons,
\begin{eqnarray}
	{\cal L} =  \frac{\eta_8}{f_8} \frac{\alpha_{\rm QED}}{4\sqrt{3}\pi}F\tilde{F} +
	\frac{\eta_0}{f_0} \frac{\alpha_{\rm QED}}{\sqrt{6}\pi}F\tilde{F} \ ,
	\label{eq:anoms}
\end{eqnarray}
where $f_8$ denotes the decay constant of $\eta_8$.
In our analysis, we adopt $\sin\theta \simeq -1/3$, $f_8\simeq 1.3\times f_\pi$, and $f_0 \simeq f_\pi$ ($f_\pi \simeq 93$\,MeV),   
which reproduce the diphoton decay widths of $\eta$ and $\eta'$~\cite{Donoghue:1986wv,Bijnens:1988kx}.
Here, we define the mixing angle between $\eta$ and $\eta'$ to be 
\begin{eqnarray}
	\eta_8 = \cos\theta\, \eta + \sin\theta\, \eta' \ , \quad
	\eta_0 = -\sin\theta\, \eta + \cos\theta\, \eta' \ .
	\label{eq:etaeta}
\end{eqnarray}

\begin{figure}[tbp]
	\centering
	\begin{minipage}{.46\linewidth}
		\includegraphics[width=\linewidth]{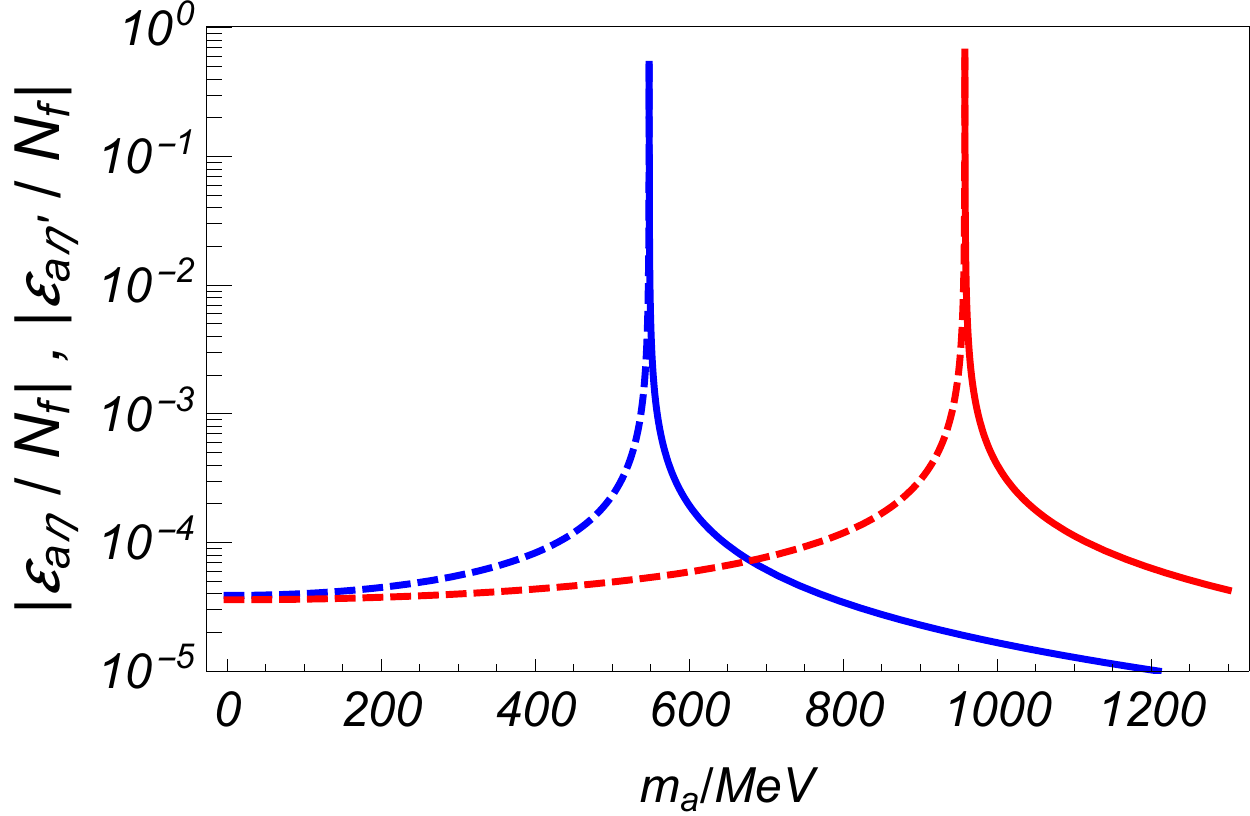}
	\end{minipage}
	\hspace{1cm}
	\begin{minipage}{.46\linewidth}
		\includegraphics[width=\linewidth]{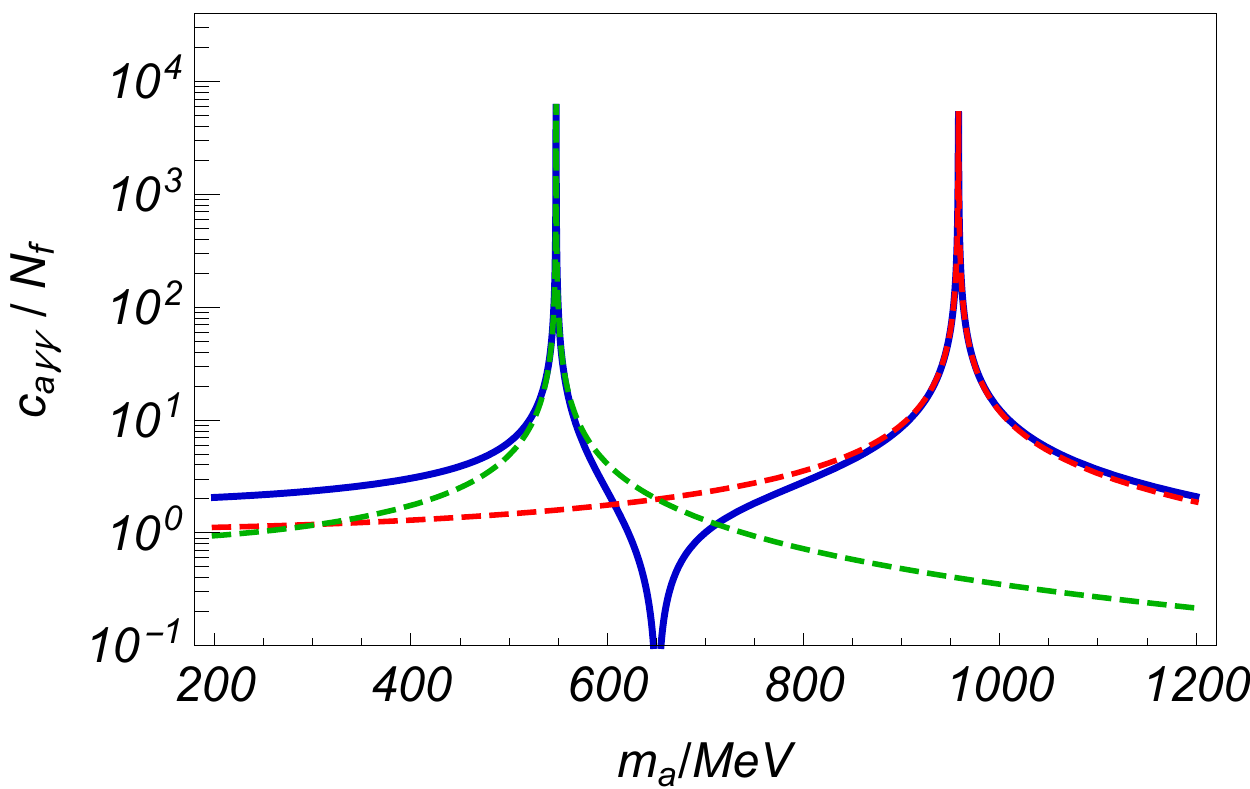}
	\end{minipage}
	\caption{\sl\small (Left) The mixing angles of the ALP to $\eta$ (blue) and $\eta'$  (red) for $f_a = 1$\,TeV. 
		The dashed curves indicate the negative mixing angles.	
		The figure shows that the mixing angles are enhanced for either $m_a \simeq m_\eta$ 
		or $m_a \simeq m_{\eta'}$.
		(Right) The effective coefficient of the anomalous coupling of the ALP in Eq.\,(\ref{eq:anom2})
		The green and red dashed lines show the contributions from the ALP mixing to $\eta$ and $\eta'$, respectively.
		The blue line shows the total contributions.
	}
	\label{fig:mixing}
\end{figure}

Now, let us estimate the mixing angles between the ALP and the $\eta$ and $\eta'$  mesons.
Through the term proportional to $m_0^2$ in Eq.\,(\ref{eq:eff0}), we obtain the mass eigenstates 
$(\eta_D, {\eta'}_D,a_D)$ 
\begin{eqnarray}
	\eta \simeq \eta_D + \varepsilon_{a\eta} a_D \ , \quad
	\eta'\simeq \eta'_D + \varepsilon_{a\eta'} a_D \ , \quad
	a \simeq a_D -  \varepsilon_{a\eta} \eta_D -  \varepsilon_{a\eta'} \eta'_D \ ,
\end{eqnarray}
with the small mixing angles  given by,
\begin{eqnarray}
	\label{eq:aetaeta}
	\varepsilon_{a\eta} \simeq 
	-\frac{N_ff_0}{{\sqrt{6}}f_a} \frac{m_{\eta'}^2}{m_{\eta}^2-m_a^2}  \sin\theta\ ,\quad
	\varepsilon_{a\eta'} \simeq 
	\frac{N_ff_0}{{\sqrt{6}}f_a} \frac{m_{\eta'}^2}{m_{\eta'}^2-m_a^2}  \cos\theta\ .
\end{eqnarray}
In the left panel of Fig.\,{\ref{fig:mixing}},  mixing angles of the ALP to $\eta$ and $\eta'$
are shown for $f_a = 1$\,TeV. 
The figure shows that the mixing angles are resonantly enhanced for either $m_a^2 \simeq m_\eta^2$ 
or $m_a^2 \simeq m_{\eta'}^2$.

%%%%%%%%%%%%%%%%%%%%%%%%%%%%%%%%%%%%%%%%%%
%%%%%%%%%%%%%%%%%%%%%%%%%%%%%%%%%%%%%%%%%%

Once the ALP gets mixed with $\eta$ and $\eta'$, a coupling constant, $c_{a{\cal O}}$, of the ALP to an operator ${\cal O}$ to which the $\eta$ and $\eta'$ mesons couple
is given by
\begin{eqnarray}
	c_{a\cal O}  = {\varepsilon}_{a\eta'} c_{\eta'\cal O} +  {\varepsilon}_{a\eta} c_{\eta\cal O} \ ,
	\label{eq:combine}
\end{eqnarray}
where $c_{\eta\cal O}$ 
and $c_{\eta'\cal O}$ are the coupling constants of $\eta$ and $\eta'$ to the
operator $\cal O$, respectively.
For example, we  obtain the effective anomalous coupling of the ALP to the photons with a normalization
\begin{eqnarray}
	{\cal L} =  
	\frac{\alpha_{\rm QED}}{4\pi}
	\frac{c_{a\gamma\gamma} }{ f_a} a
	F\tilde{F} \ ,
	\label{eq:anom2}
\end{eqnarray}
as shown in the right panel of Fig.\,\ref{fig:mixing}.
The suppression of $c_{a\gamma\gamma}$ at around $m_a \simeq 700$\,MeV 
is caused by a destructive interference between the contributions of $\eta$ and $\eta'$.

\begin{figure}[t]
	\begin{center}
		\begin{minipage}{.46\linewidth}
			\includegraphics[width=\linewidth]{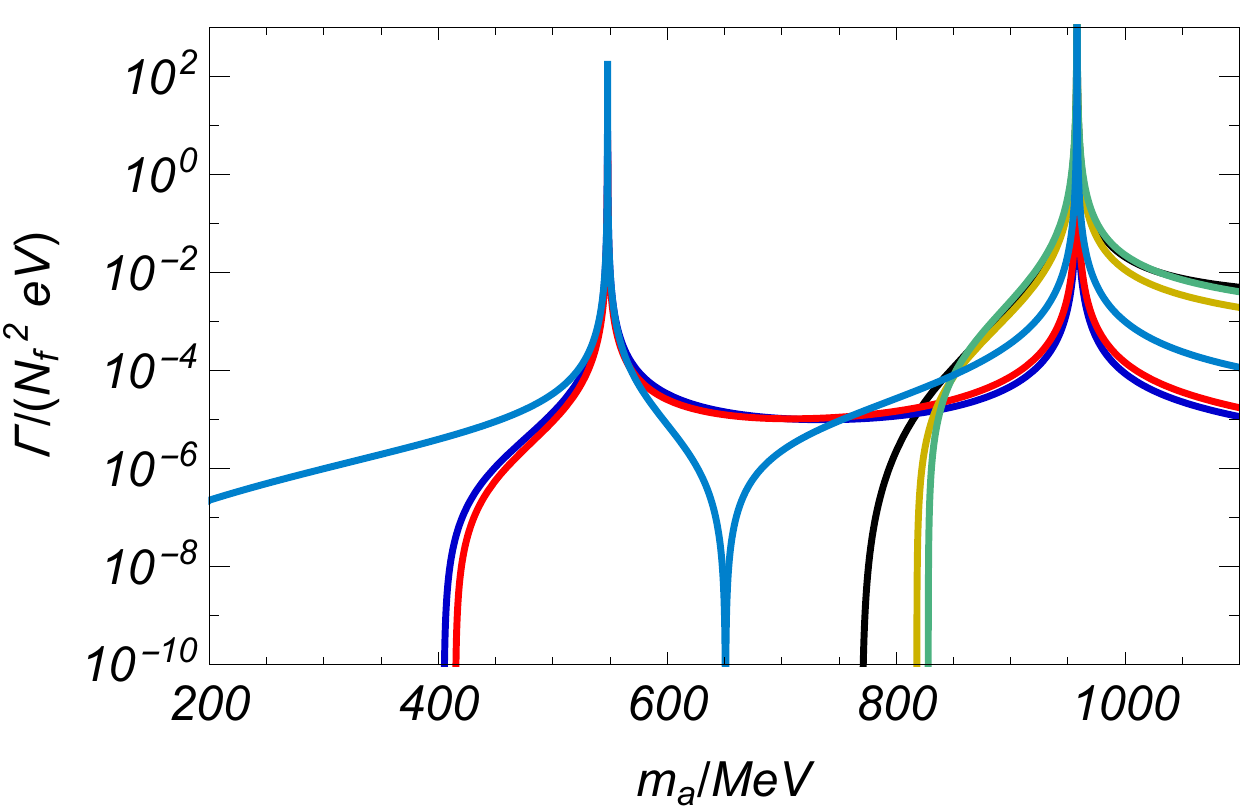}
		\end{minipage}
		\hspace{1cm}
		\begin{minipage}{.46\linewidth}
			\includegraphics[width=\linewidth]{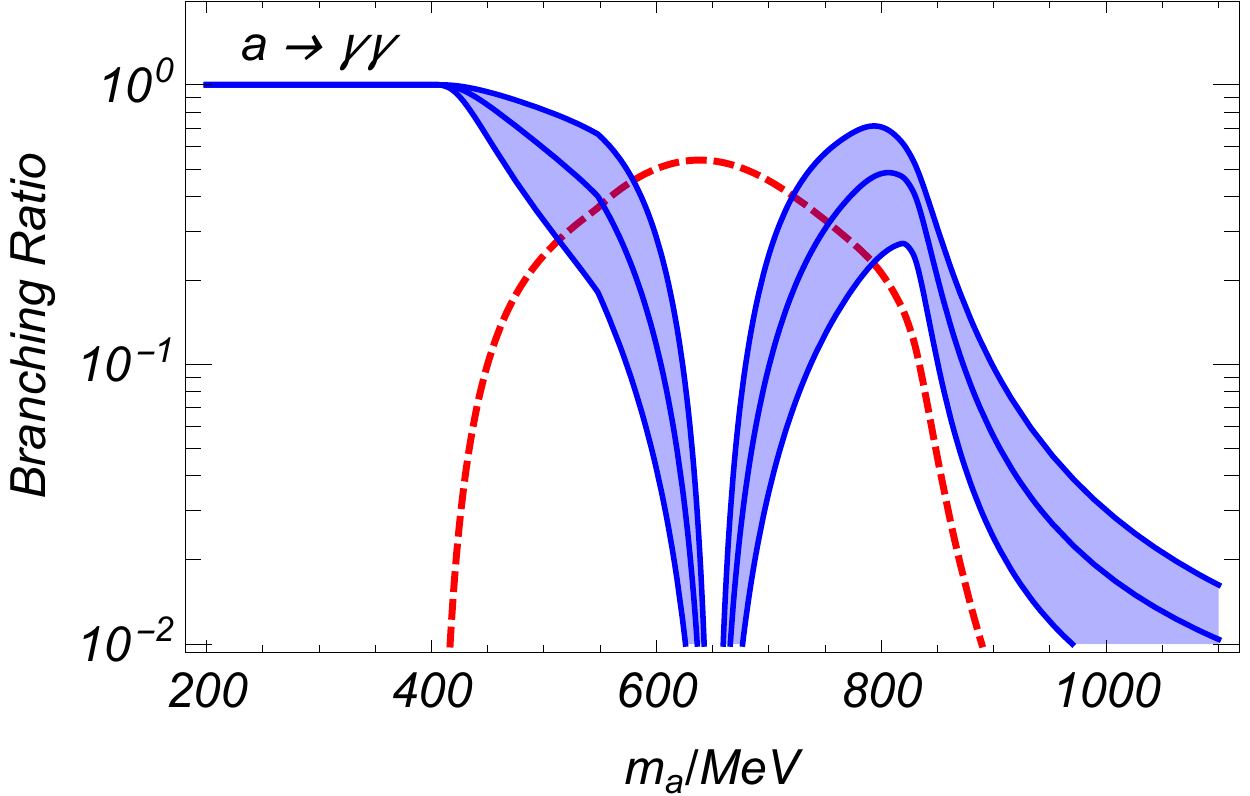}
		\end{minipage}
	\end{center}
	\caption{\sl \small
		(Left) The relevant partial decay widths of the ALP into $\gamma\gamma$ (light blue), 
		$3\pi^0$ (dark blue), $\pi^0\pi^+\pi^-$ (red), 
		$\rho+\gamma$ (black), $\eta +2\pi^0$ (grass green),
		$\eta + \pi^+ + \pi^-$ (green) for $f_a = 1$\,TeV.
		(Right)
		The branching ratio of the ALP into $2\gamma$ for $f_a = 1$\,TeV (blue band).
		The  band represents ${\cal O}(1)$ uncertainties of our estimations 
		of the three body decay modes from the observed decay widths of $\eta$ and $\eta'$
		(see Ref.~\cite{Chiang:2016eav}).
		We also show the branching ratio into $3\pi_0$ which also leads to the photon jet signal,
		although it has ${\cal O}(1)$ uncertainty.
	}
	\label{fig:width}
\end{figure}

By knowing the mixing angles, the decay widths of the light ALP can be  estimated from the decay widths
of $\eta$ and $\eta'$ mesons (see e.g.~\cite{Agashe:2014kda}).
In the left panel of Fig.\,\ref{fig:width}, we show the estimated partial decay widths of the ALP.
We also show the branching ratios into $2\gamma$ and $3\pi_0$ in the right panel.
In our estimation, we approximate the amplitudes of the three body decay modes of $\eta$ and $\eta'$
by the square root of the decay widths divided by the phase space volume.
Then, we combine them constructively by using Eq.\,(\ref{eq:combine}).
Accordingly, our estimates of the decay widths into three body modes have ${\cal O}(1)$ uncertainties.
For more details, see the appendix of Ref.~\cite{Chiang:2016eav}.

%%%%%%%%%%%%%%%%%%%%%%%%%%%%
%%%%%%%%%%%%%%%%%%%%%%%%%%%

The right-panel of the figure shows that $Br(a\to 2\gamma) = {\cal O}(1)$ is achieved for a wide range 
of parameter space.%
\footnote{For multiple vector-like colored fermions, the production 
	cross section in Eq.\,(\ref{eq:production}) is enhanced, and hence, the signal cross section
	can be achieved even for a slightly smaller $Br(a\to 2\gamma)$.
}
Thus, this simple model based on  chiral symmetry breaking of the vector-like fermion
provides a very good example of the models where the $750$\,GeV resonance decays 
into a pair of photon-jets with $N_\gamma = 2$.
It should be also noted that, in this model, the ALP also decays into other light hadrons.
In particular, the ALP decays into $3\pi_0$ which subsequently decay into $6\gamma$.
Therefore, this model also provides an example of $N_\gamma = 6$.

In this example, the photon-jets with $N_\gamma = 4$ are not expected, since 
the ALP decays into photons only thorough the mixing with $\eta$ and $\eta'$.
In other class of the models, however,  it is also possible to have models with $N_\gamma = 4$ 
in e.g. Ref.~\cite{Chang:2015sdy} in which the photon-jet is made by a decay of a light 
$CP$-even scalar particle decaying into a pair of neutral pions.
Thus, in the following study, we also take into account the case with $N_\gamma = 4$.
As we will show, the photon-jets for $N_\gamma \ge 2$ can be immediately distinguished by using
the distribution of the sum of $p_T$ of the first $e^+e^-$ pair from the photon conversion.

\begin{figure}[t]
	\begin{center}
		\begin{minipage}{.46\linewidth}
			\includegraphics[width=\linewidth]{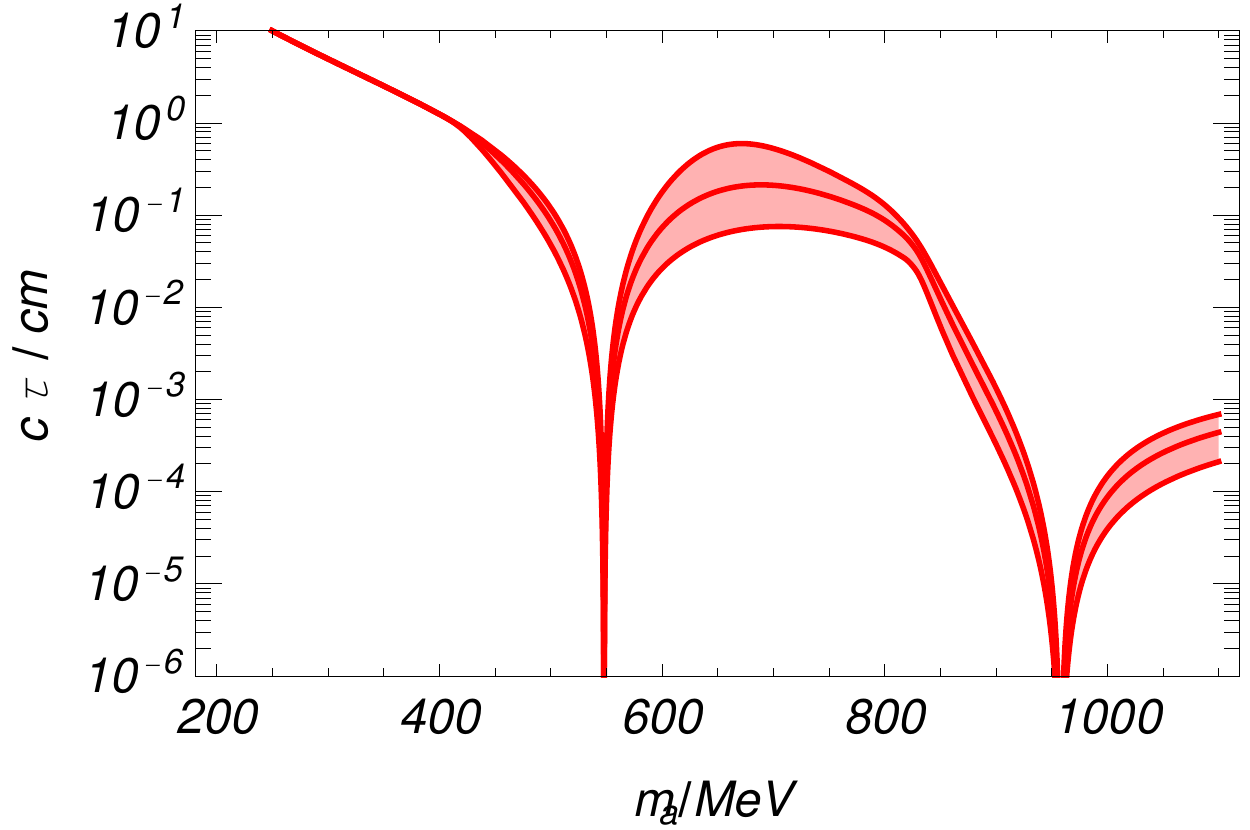}
		\end{minipage}
		\hspace{1cm}
		\begin{minipage}{.46\linewidth}
			\includegraphics[width=\linewidth]{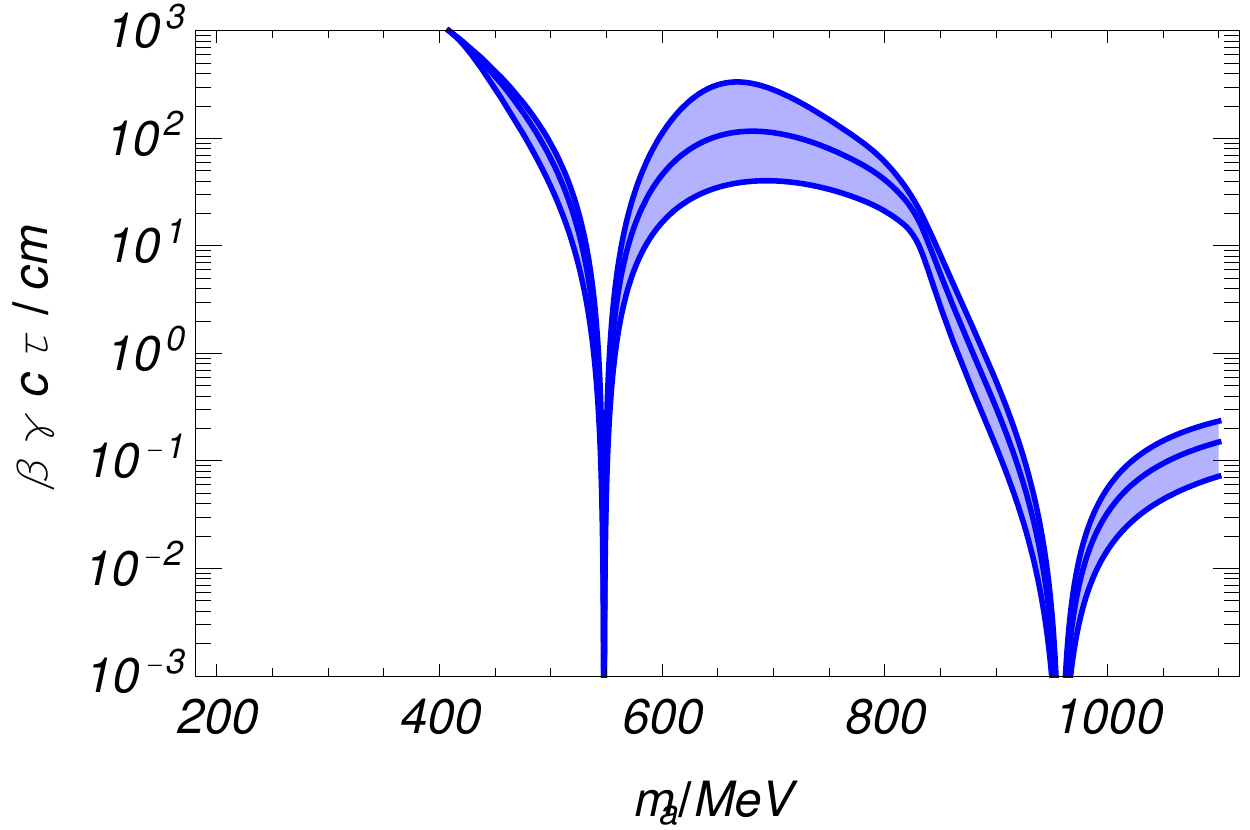}
		\end{minipage}
	\end{center}
	\caption{\sl \small
		(Left)
		The ALP lifetime multiplied by the speed of light for $f_a =1$\,TeV and $N_f=2$.
		(Right)
		The boosted decay length of the ALP for $f_a = 1$\,TeV and $N_f=2$
		when the ALP is produced by the two-body decay of the $750$\,GeV resonance 
		at rest (i.e. $\gamma \simeq 325\,{\rm GeV}/m_a$ with $\beta =1/ \sqrt{1-\gamma^2}$).
		In both panels, the bands represent ${\cal O}(1)$ uncertainties of our estimations 
		of the three body decay modes. 
	}
	\label{fig:length}
\end{figure}

Before closing this section, let us show the decay length of the ALP in Fig.\,\ref{fig:length} for $f_a = 1$\,TeV and $N_f=2$. Here, we need $N_f > 1$ to make the lifetime of the ALP short enough so that the majority of ALP decay inside the inner tracker.
In the following, we fix $m_a=400\,\text{MeV}$ and assume $N_f>1$. As is discussed in \cite{Chiang:2016eav}, large $N_f$ is motivated because, in that model, we need to  cancel the coupling between the ALP and hidden photons to suppress entropy in the hidden sector in the early Universe.

%%%%%%%%%%%%%%%%%%%%%%%%%%%%
%%%%%%%%%%%%%%%%%%%%%%%%%%%

%%%%%%%%%%%%%%%%%%%%%%%%%%%%%%%%%%%%%%%%%%%%%%%%%%%%%%
\section{Photon-jets in ATLAS and CMS detectors}
\label{sec:detector}
\begin{figure}
\includegraphics[width=.65\linewidth]{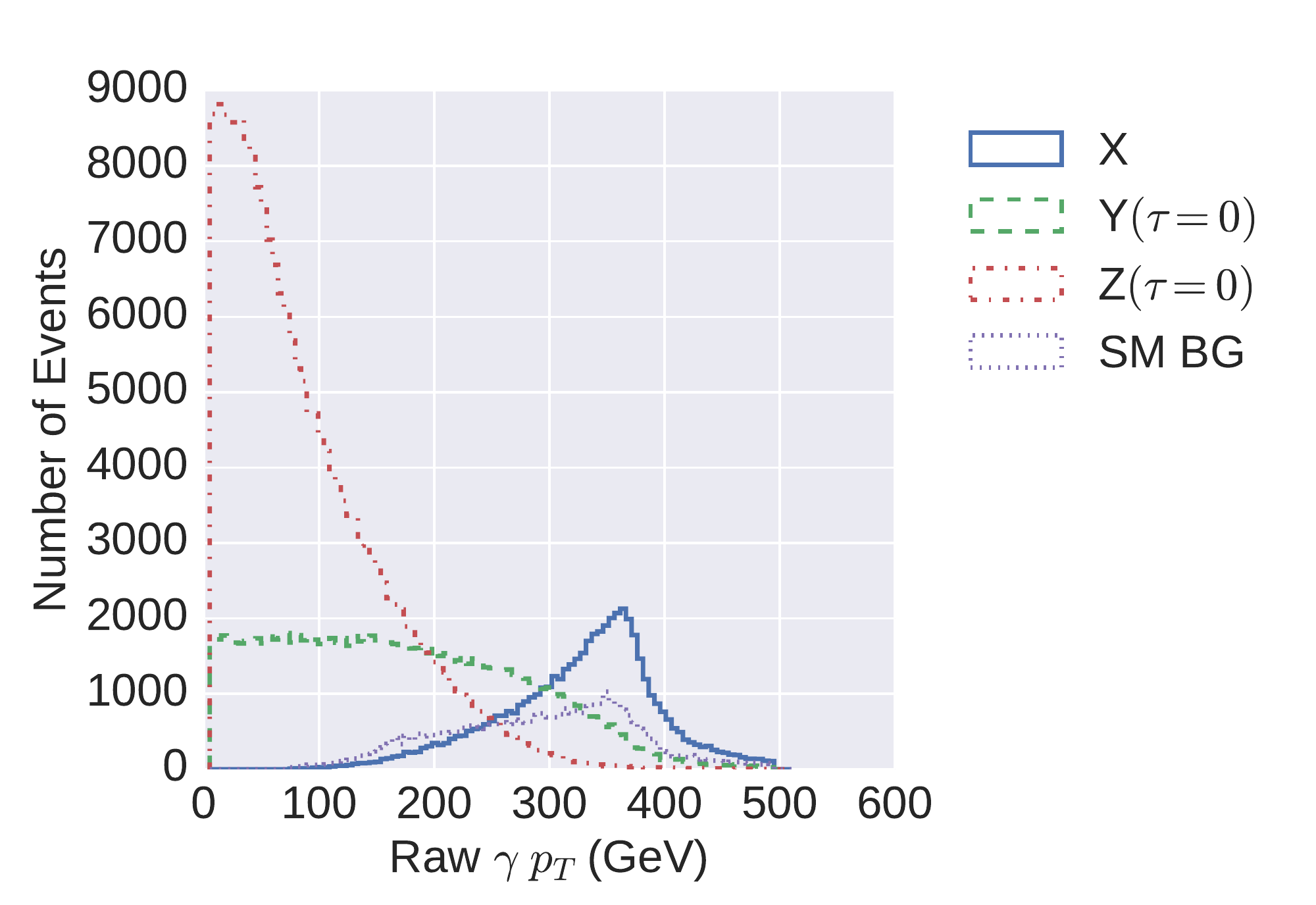}
\caption{\sl\small The $p_T$ distribution of photons in the diphoton resonance (X), the di-photon-jet resonance with $N_\gamma = 2$ (Y) 
and, the one with $N_\gamma=4$ (Z), respectively.  
We also show the SM background in 700\,GeV$<m_{\gamma\gamma}<$800\,GeV. 
Here, we assume that the intermediate particle decays immediately after its production, i.e. $\tau = 0$. }
\label{fig:photonpt}
\end{figure}

In this section, we describe photon-jet signature in  ATLAS and CMS detectors with particular attentions to photon conversion. 
We first show  photon $p_T$  distribution inside  the photon-jet for $N_{\gamma}=2$ and $4$ in Fig.\,\ref{fig:photonpt}.    
Here, we generate $40000$ events of the resonance production, $pp \rightarrow s(750\,{\rm GeV})$, 
through the effective coupling  ${\cal L} \propto  s G_{\mu\nu} G^{\mu{\nu} }$ in Eq.\,(\ref{eq:eff1}) 
using Feynrule \cite{Alloul:2013bka}  and Madgraph \cite{Alwall:2014hca}.
The resonance $s$ subsequently decays into $s\rightarrow \gamma \gamma$,
$s\rightarrow a a \rightarrow  (2\gamma)\  (2\gamma)$,  or $s \rightarrow a a \rightarrow (2\pi^0)\   (2\pi^0) \rightarrow (4\gamma)   (4\gamma)$ 
assuming spherical distribution. 
Hereafter, we name these signal models X ($N_\gamma =1$), Y ($N_\gamma = 2$), and Z ($N_\gamma =4$), respectively.   
In Fig.\,\ref{fig:photonpt}, we also show the SM background distribution of $p p \rightarrow \gamma \gamma$ + up to $ 2$-jet processes generated 
by Madgraph between 700\,GeV$<m_{\gamma\gamma}<$800\,GeV for comparison.
The SM background processes are generated as many as $40000$ events.
In the figure and throughout this paper, we require the following cuts as in \cite{CMS:2015dxe}:
\begin{itemize}
	\item At least, one pseudorapidity of the photon(-jet)s must be within the barrel region, $|\eta| < 1.44$.
	\item $|\eta|$ of the other photon-jet must be less than $1.869$.
	\item The photon(-jet)s in the region $1.44 < |\eta| < 1.57$ are dropped.
	\item The sum of $p_T$ of photons in a jet are required to be greater than $75\,\text{GeV}$.
\end{itemize}

As a rule of sum, each photon-jet tends to have $p_T^{\gamma{\rm -jet}}=\sum_{i=1}^{N_{\gamma}}p_T^{i}\sim m_s/2$~GeV, 
therefore the constituent photons have $p_T^{i}$ of around $p_T^{\gamma{\rm -jet}}/N_{\gamma}$ in average.  
As long as $m_a\ll p_T^{\gamma{\rm -jet}}$, those photons are 
very collinear and  they cannot be separately measured   by ECAL only. The typical separation between photons in our setup is $\Delta \eta \sim \Delta\phi\sim400\,\text{MeV}/375\,\text{GeV} < 0.001$ while the ECAL position resolution is $\mathcal O(\pi/180)\sim\mathcal O(0.01)$\,\cite{Bayatian:2006zz}.

%%%%%%%%%%%%%%%%%%%%%%%%%%%%%%%%%%%%%%%%%%%%%%%%%%%%%%
%%%%%%%%%%%%%%%%%%%%%%%%%%%%%%%%%%%%%%%%%%%%%%%%%%%%%%

Now, let us discuss how the photon-jets interact with  
the detector material more closely.
Some of the photons in the photon-jets
convert into $e^+e^-$ pairs when they go through the detector material.  
The produced  $e^{\pm} $ subsequently loses its energy by the bremsstrahlung.  
Typical amount of matter required for the photon conversion or the bremsstrahlung is given by the radiation length  $X_0$. 
The radiation length corresponds to either (a) the mean distance which electron energy $E_e$ becomes $ E_e/e$ by bremsstrahlung   
or (b) $7/9$ of the mean free path for $e^+e^-$ pair production by a high-energy photon.  
Both ATLAS and CMS  have inner trackers  with the thickness corresponding to about $0.4 X_0$ at $\eta=0$ 
and to about $2X_0 $ in the forward region\,\cite{Chatrchyan:2014fea,Aad:2008zzm}.  After passing through the inner trackers, the electrons 
and photons develop electromagnetic showers and its total energy are measured by the ECAL.
Photons are classified as unconverted photons if their ECAL activities do not match with
the track or reconstructed conversion vertex in the inner detector. Those with a matching
reconstructed conversion are classified as converted photons. 
   
 The number of photons in the photon-jets may be studied by looking for anomalies in the electromagnetic shower profile.  
 In particular, the fraction of the converted photon to the identified photon at the ECAL 
 and the energy  fraction of  the $e^+e^-$ pair  to the photon-jet energy measured at the ECAL contain important information. 
 The fraction, $P^{\gamma-{\rm jet}}_{\rm conv}$, for example, must increase with $N_{\gamma}$, 
 \begin{equation}
P^{\gamma-{\rm jet}}_{\rm conv}= 1-(1-P^{\gamma}_{\rm conv})^{N_{\gamma}}\ .
 \end{equation}
Here, $P^{\gamma}_{\rm conv}$ is a single photon conversion probability, 
  \begin{equation}
 P^{\gamma}_{\rm conv}= 1- \exp\left[-\frac{7X}{9X_0}\right] 
 \end{equation}
with $X$ being the length in the unit of $[M]/[L^2]$ through which the photon  has passed. 
 The energy of $e^+ e^-$ pair $E_{\rm pair}$, if it can be measured,  would be significantly smaller  than $ E_{\gamma-{\rm  jet}}$, roughly 
 $ E_{\gamma-\text{jet}}/(2N_{\gamma})$. 
 The distribution of  $e^+$ or $e^-$ energy from the conversion vertex is given by 
 \begin{equation}\label{conversion}
 \frac{d\sigma}{dx}\propto 1-\frac{4}{3}x(1-x)
 \end{equation}
where  $x=E_e/E_{\gamma}$\,\cite{Tsai:1973py,Agashe:2014kda}. 

In addition, the lifetime of the intermediate particle $a$ may be measured by looking for deficiency of the photon conversion 
within the boosted decay length,  $\beta\gamma c \tau_a$, from the beam collision point,  since conversion occurs only after
the photons being produced by the decay of $a$.
Therefore, reconstruction of the distance between interaction point and the conversion position  $L_{\rm conv}$ is also important for our purpose.  

The fraction of the converted photons to the isolated single high $p_T$ photons identified at the ECAL are reported by ATLAS
\cite{EGAM2015004} and CMS \cite{Khachatryan:2015iwa}.  
According to those studies, more than 20\%\,(40\%) of the photons 
are  converted  at $\eta=0\,(1)$ in both ATLAS and CMS.  
The resolution of the converted photon momentum is also compared with the Monte Carlo (MC) simulation for both ATLAS and CMS in  
the $Z\rightarrow \mu\mu\gamma$ mode with conversion\,\cite{ATLAS-CONF-2012-123,Khachatryan:2015iwa}.  
The $\gamma$ resolution is obtained by using both track hits and calorimeter information matched to the track. 

For the photon-jet signal, on the other hand, the energy fraction of the produced  $e^{\pm}$ at the photon conversion 
should be much smaller than the one expected for the single photon.
Therefore, the momentum measurement of the converted photon solely by the track measurements is crucial to distinguish  photon-jets 
from  single photons.
A track of a charged particle in the uniform magnetic field  $B$ ($B\sim 2$\,T for ATLAS and $\sim 4$\,T  for  CMS for the inner trackers) 
is measured by utilizing the hits in the tracker.  
In three-point measurement, the relation between $p_T$\,[GeV] and  sagitta $s$\,[m] for the radial chord length  $L$\,[m] from the origin 
of the track is\,\footnote{For multi-point measurement, the simple equation, Eq.\,\ref{sagitta}, does not hold.}
\begin{equation}\label{sagitta}
p_T \sim 0.3\times \frac{L^2 B[{\rm T}] }{8\,s}\ ,
\end{equation}
therefore 
\begin{equation}\label{sagittae}
\frac{\delta p_T}{p_T}= \frac{8\,p_T\,\Delta s}{0.3\, L^2B}\ . 
 \end{equation}
 The position resolutions of the CMS sensor layers are 
 typically  20--40 $\mu m$. Putting this resolution  into Eq. (\ref{sagittae}), 
 we obtain $\delta p_T/p_T\sim 1.4$--$3 $\% for $L=1\,$m and $p_T = 100$\,GeV.  
% ATLAS  PIXEL and SCT 
% position resolutions are similar $\sim O(10)$$\mu$m , while  the TRT  placed  at radial distance from the beam $55$~cm $<R<108$~cm has significantly worse resolution  of $\sim170\mu m$, but provide continuous track monitoring.   
The resolution of the track momentum at $p_T=100$\,GeV for  isolated $\mu$, $\pi^\pm$, and $e^{\pm}$ using tracker information
are documented in detail for CMS\,\cite{Chatrchyan:2014fea}. 
There, the $p_T$ resolution is less than 2\% at $p_T=100$\,GeV for a single isolated $\mu^\pm$ and $\pi^\pm$  in the central region,
which is consistent with the rough estimation based on Eq.\,(\ref{sagittae}).  
It should be noted that  Eq.\,(\ref{sagittae}) shows that the resolution becomes worse for a larger $p_T$ or a smaller  $L$.
Thus, when the conversion takes place at a outer layer of the trackers, the effective $L$ is small. 
For example, if the track starts at a radial position $R$, $L=R_{\rm max}-R$  must be used, 
where $R_{\rm max}$ is the radial size of the inner tracker system, $R_{\rm max}\simeq 1$\,m, for both ATLAS and CMS. 

One has to keep in mind that measuring $e^{\pm}$ $p_T$ is not straightforward task compared with those of $\mu^\pm$ or $\pi^\pm$, 
because the electron keeps losing its energy by bremsstrahlung.  
Accordingly, the resolution of the electron momentum  is  about 15\% for  $p_T=100$\,GeV, which is a factor of 7 worse compared with
those of $\mu^\pm$ and $\pi^{\pm}$. 
In short, once a photon converts to $e^+e^-$,  
 many photons  going in  the same directions 
 are also produced, so that  $e^{\pm}$ tracks are 
 mis-measured. 
 The effect can be partially compensated by improving track reconstruction algorithm for an electron candidate. 
 However, the distribution of the ratio between the electron energy measured by the calorimeter\,\footnote{Electron energies measured in ECAL is much more accurate than in the tracker.} to the electron momentum  reconstructed from the track hits, 
 $E/p_{\rm track}$,  has a long tail in $E/p_{\rm track}>1$, and the events in the tail arise due to the bremsstrahlung involving relatively hard photon emissions.
The cross section of bremsstrahlung is given in 
Ref.\,\cite{Agashe:2014kda} as 
\begin{equation}\label{brem}
\frac{d\sigma}{dk}= \frac{A}{X_0N_A k}\left( \frac{4}{3}-\frac{4}{3}y+y^2\right),
\end{equation}
where $k$ is energy of photon emitted from an electron by bremsstrahlung, and 
$y=k/E_e$, $N_A$ is Avogadoro's number and $A$ is the atomic mass of the absorber.   
The formula is a good approximation for an electron with energy $10$\,GeV$<E_e <1$\,TeV except for a small $y$ region.  
The rate of energy loss by the photons with momentum fraction between  $y$ and $y+dy$, i.e. $y d\sigma/ dy $, is  a rather flat function, 
 and easily implemented in MC simulation.

 % In the next section, we try to study the LHC sensitivity to the 
% photon jet with special attention to events with converted photons. 
 %We will see that the background photon 
% contamination reduces significantly by  measurement of the track 
%$p_T$ associated  with the photon candidate.  The electron $p_T$ %measurement also improve the measurement of the number 
% of photon in the photon jet. We will also see that the LHC 
 %experiment is sensitive to the life time of the particle $a$. 

% At the end of this section, we summarize our assumptions of the detector performance used in the analysis in the next  section. 
In the following, we estimate the LHC  sensitivities  based on the geometry of 
the CMS detector for simplicity, which has inner trackers  consisting of  
the silicon sensors  with uniform position sensitivity.  
In the case of the ATLAS detector, on the other hand, its outer half of the inner detector
consists  of the TRT, whose position resolution is rather restrictive compared with the inner pixel and strip sensors and detailed detector simulations are necessary to estimate the performance.% 

%%%%%%%%%%%%%%%%%%%%%%%%%%%%%%%%%%%%%%%%%%
\begin{figure}[tbp]
	\centering
  \includegraphics[width=.6\linewidth]{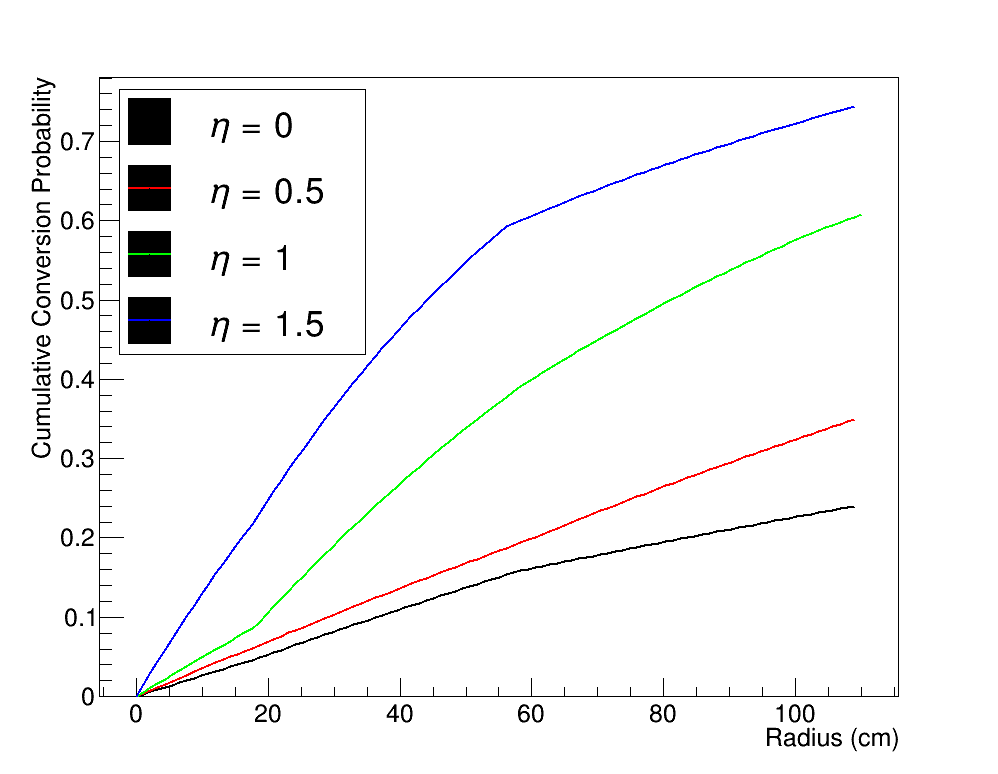}
	\caption{\sl\small Cumulative photon conversion probability for given $\eta$ for the simplified CMS detector used in this analysis.
	The geometry and material distribution of the CMS detector are taken from \cite{Chatrchyan:2014fea}.
	}
	\label{fig:cumulative}
\end{figure}
%%%%%%%%%%%%%%%%%%%%%%%%%%%%%%%%%%%%%%%%%%

\begin{figure}[tbp]
	\centering
  \includegraphics[width=.65\linewidth]{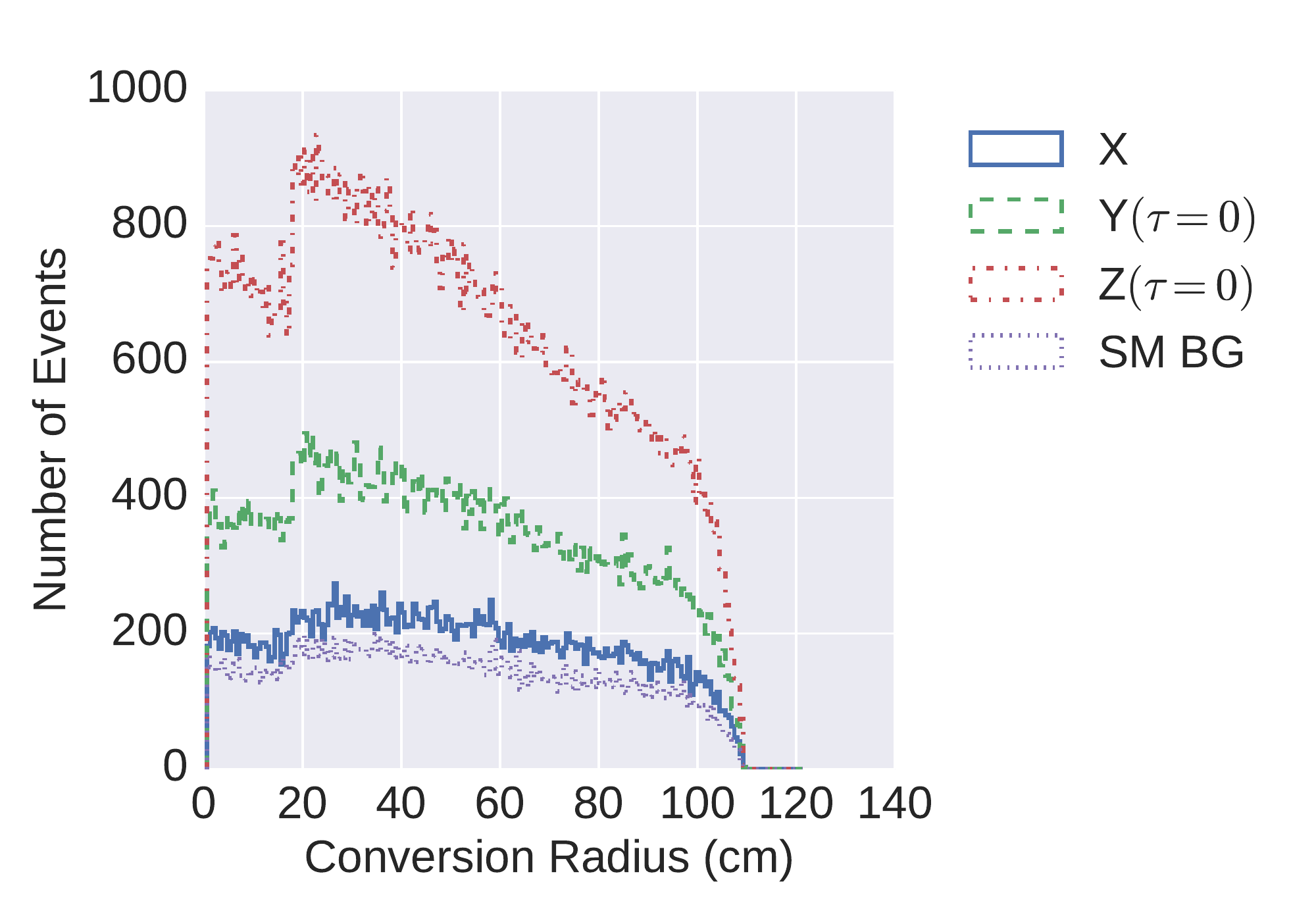}
	\caption{\sl\small Distribution of the radial position of the photon conversion. 
	The labels are the same as those in Fig.\,\ref{fig:photonpt}.}
	\label{fig:conversionpoint}
\end{figure}

In our MC simulation, we consider the above mentioned three signal models (X) $ s\rightarrow \gamma\gamma$,  
(Y) $s\rightarrow a a \rightarrow  (2\gamma)\  (2\gamma)$, and  (Z) $s \rightarrow a a \rightarrow (2\pi^0)\   (2\pi^0) \rightarrow (4\gamma)\   (4\gamma)$.
Then, each $\gamma$ converts to an $e^+ e^-$ pair according to the conversion probability estimated from
the radiation length of each detector components for given $\eta$ as shown in Fig.\,2 of \cite{Chatrchyan:2014fea}.  We read the radiation length  of each detector component for sufficintly many $\eta$ values and interpolate them for a generic $\eta$.
Only the $e^+e^-$ pair of first converted photon is used in our analysis, because the other converted 
photons overlap with photons from bremsstrahlung of the first converted photon. 
The geometry of the CMS detector is taken from Fig.\,1  of \cite{Chatrchyan:2014fea}. 
Although the CMS detector consists of the strip detector layers separated by ${\cal O}(10)$\,cm, 
we use averaged (and continuous) material distribution of each detector component  to estimate  position of conversion. 
We show the cumulative photon conversion probability used in our analysis in Fig.\,\ref{fig:cumulative} for some representative value of $\eta$. 
The conversion position is also shown in Fig.\,\ref{fig:conversionpoint}. We cannot compare it with CMS data as no similar plot is available. In the case of ATLAS detector, the conversion probability is consistent with the radiation length of the detector components\,\cite{Andreazza:2009zz}.

%%%%%%%%%%%%%%%%%%%%%%%%%%%%%%%%%%%%%%%%%%
%%%%%%%%%%%%%%%%%%%%%%%%%%%%%%%%%%%%%%%%%%

\begin{figure}[tbp]
\includegraphics[width=.65\linewidth]{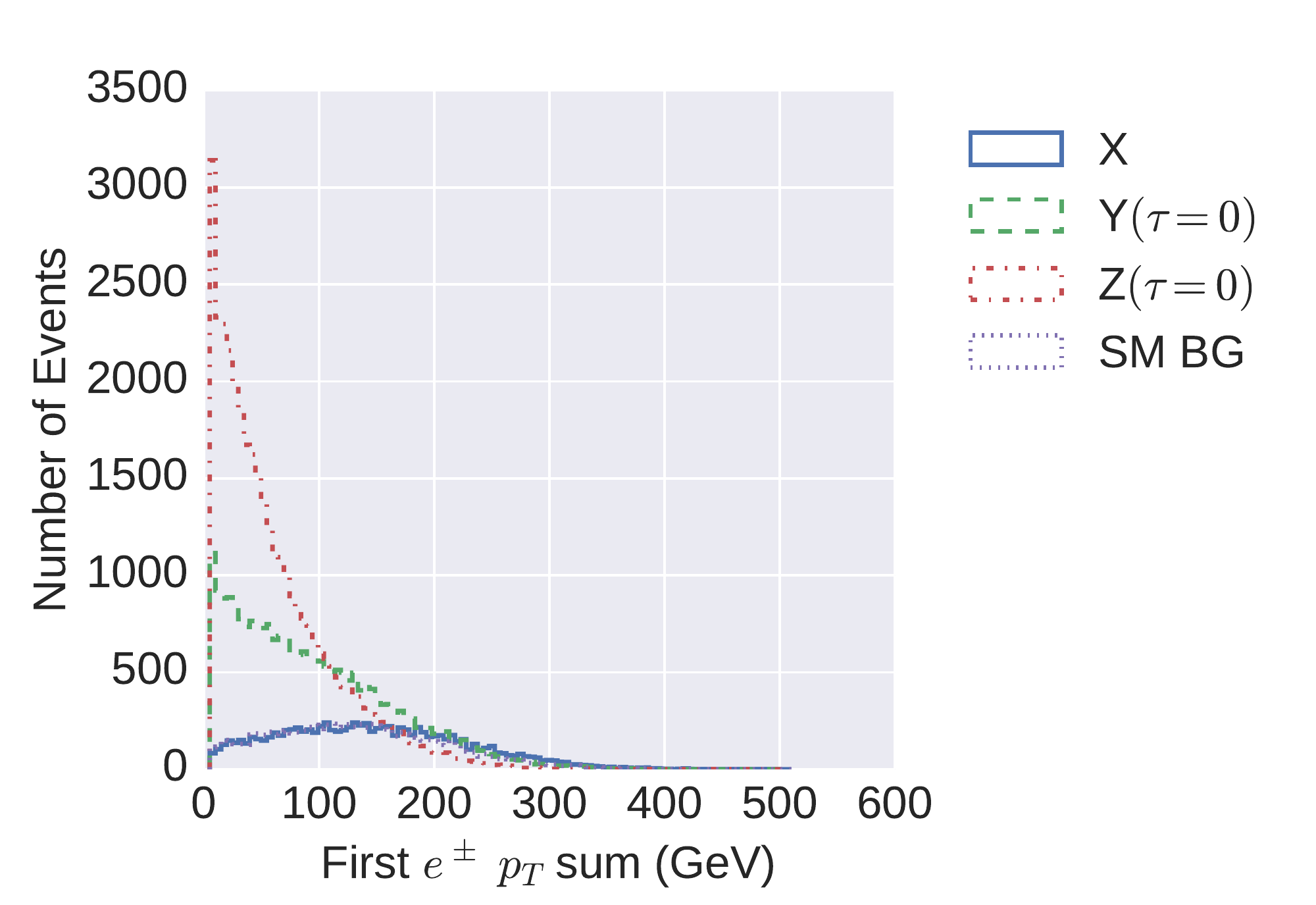}
\caption{\sl\small The smeared distributions of the sum of the $p_T$ of the first $e^+e^-$ pair
measured by tackers.
Events are removed if the error of $e^+e^-$ momentum is more than 30\%. 
The labels are the same as those in Fig.\,\ref{fig:photonpt}.
} \label{fig:epairpt}
\end{figure}

The $e^+$ and $e^-$ momentum  distributions  are estimated by using collinear  splitting with the probability given in Eq.\,(\ref{brem}).  
Because  bremsstrahlung after the conversion is essential for the tail distribution of $E_e/p_{\rm track}$, we estimate  the photon emission from 
bremsstrahlung  for the photon with momentum $k$ with $k> k_{cut}=\max(5\,\text{GeV}, 0.1 E_e)$ using Eq.\,(\ref{brem}) by Monte Carlo simulation.
We record the electron and positron momenta when $e^+e^-$ tracks are separated by at least $50\,\mu\text{m}$ and they have passed at least three layers of the trackers.
We assume the smearing due to photon emission below $k_{cut}$ will be taken care by electron tracking algorithm, 
so that we can approximate it by  gaussian smearing consistent with 68\% error given  in \cite{Agashe:2014kda}. 
From the lower-right panel of Fig.\,17 in \cite{Chatrchyan:2014fea}, we take $1\sigma$ resolution  of electron momentum as 15\% at $p_T=100$\,GeV as a canonical value, then scale it depending on the conversion position and the momentum.   
Namely, based on the above discussions, we scale  electron momentum resolution $\delta p_T/p_T$ linear in $p_T$ 
but not better than 6\% which is the resolution of electron  momentum at $p_T=10$\,GeV.  
%Below that scale, the electron momentum resolution is taken to be constant as indicated in the study.
 We also scale the momentum resolution depending on the conversion radius  $R_\text{conv}$, that is, 
\begin{equation}
\frac{\delta p_T}{p_T}={\rm max}\left [6\%, \ 15\%\times \frac{p_T}{100{\rm GeV}}\right]\times \frac{R^2_{\rm max}}
{(R_{\rm max}-R_{\rm conv})^2}.  
\end{equation}
If $\delta p_T/p_T>30$\%, we regard the momentum cannot 
be reliably measured although the track is identified.  This means 
we do not use momentum information for the events with conversion radius $R> 0.7R_{\rm max}$ in our estimate. In the ATLAS detector, the half of the tracker consists of TRT. If we do not use the information from TRT, the effective region is therefore less than half of its full radius. This is not very different from CMS and a similar analysis to ours may be performed in the ATLAS detector as well. Finally, we 
ignore the efficiency of track measurement and 
photon identification. It is typically 90\% for 
the range of momentum we are interested. 
 We assume track with $p_T<5$\,GeV
is dropped. 
The smeared momentum distribution of the first $e^+e^-$ pairs for each resonance models, (X), (Y), (Z),
and the SM background are shown in Fig.\,\ref{fig:epairpt}.

So far, we have included the effect of the bremsstrahlung and the smearing. In particular, inclusion of bremsstrahlung effect is necessary to reproduce a tail of $E/p_\text{track}$ distributions in large $E/p_T$ region. Such a tail plays an important role to estimate the standard model background contamination in low $p_\text{track}$ region in the next section. We believe that our simulation is qualitatively acceptable, as the distribution at $Z\to\mu\mu\gamma$ measurement\,\cite{Khachatryan:2015iwa} indeed have such a tail.

%%%%%%%%%%%%%%%%%%%%%%%%%%%%%%%%%%%%%%%%%%%%%%%%%%%%%%%%
%%%%%%%%%%%%%%%%%%%%%%%%%%%%%%%%%%%%%%%%%%%%%%%%%%%%%%%%

%%%%%%%%%%%%%%%%%%%%%%%%%%%%%%%%%%%%%%%%%%%%%%%%%%%%%%%%
\section{Analysis using $p_T$ of converted photon tracks}
\label{sec:photonjets}
In this section, we show how well the di-photon-jets resonance can be distinguished from 
the diphoton resonance by using $p_T$ information of converted photons. 
As  introduced in the previous section, we consider three signal models, (X) $N_\gamma = 1$, (Y) $N_\gamma= 2$, and (Z) $N_\gamma = 4$.

%We study the distribution of the $p_T$ sum of the first $e^+e^-$ pair from photon conversion for both signal and background. 
The distributions of the $p_T$ sum of the first $e^+e^-$ pair from photon conversion, $p_T^{\rm sum}$,
are determined by generating $40000$ events for each model and the SM background, so that the statistical errors of the distributions are negligible.
The results have been shown in Fig.\,\ref{fig:epairpt}. 
In the smeared distributions, events containing $e^\pm$ with $p_T\gtrsim200\,$GeV are removed due to poor momentum resolution. 
The figure shows that the SM background events as well as the events of the model X can be efficiently reduced by putting an upper limit on  $p_T^{\rm sum}$.

%%%%%%%%%%%%%%%%%%%%%%%%%%%%%%%%%%%%%%%%%%
\begin{table}[tbp]
        \centering
        \caption{\sl\small The fractions of event numbers in each category for given models.
        }
\begin{longtable}[]{@{}c|cccccc@{}}
\toprule
& A & B & C & D & E & F\tabularnewline
\midrule\hline
\endhead
X & 0.272 & 0.022 & 0.035 & 0.040 & 0.084 & 0.547\tabularnewline
Y\((\tau = 0)\) & 0.222 & 0.149 & 0.121 & 0.086 & 0.093 &
0.329\tabularnewline
Z\((\tau = 0)\) & 0.186 & 0.363 & 0.182 & 0.082 & 0.047 &
0.140\tabularnewline
SM BG & 0.244 & 0.036 & 0.055 & 0.062 & 0.102 & 0.500\tabularnewline
\bottomrule
\end{longtable}
        \label{tab:binDist}
\end{table}
%%%%%%%%%%%%%%%%%%%%%%%%%%%%%%%%%%%%%%%%%%

In our analysis, we classify the events into six categories;
\begin{description}
\item[A] Events with $\delta p_T/p_T > 30$\% for at least one of the converted electrons.
\item[B] Events with $p_T^\text{sum} \le 50\,\text{GeV}$.
\item[C] Events with $50\,\text{GeV} < p_T^\text{sum} \le 100\,\text{GeV}$.
\item[D] Events with $100\,\text{GeV} < p_T^\text{sum} \le 150\,\text{GeV}$.
\item[E] Events with $150\,\text{GeV} < p_T^\text{sum}$.
\item[F]  Events with no photon conversion by the end of the tracker.
\end{description}
Fractions of the events falling in each category  estimated from the MC samples are given in Tab.\,\ref{tab:binDist}.

Now, let us assume that one of the models, X, Y, or Z  is the true model, whose signal cross section after the cuts given in Sec.\,\ref{sec:detector} is $s_0$.\footnote{The cross section before the cut is almost twice as large as $s_0$.}
The MC samples of the true model are generated for a given integrated luminosity $\mathcal L_0$.
The background samples are also generated where the SM background cross section after the cuts is fixed to $1\,\text{fb}$. This value is estimated from \cite{ATLAS:2015aa,CMS:2015dxe, collaboration:2016aa,CMS:2016owr}.
The signal events are then fit to a mixed distribution of the two of the models X, Y and Z and the SM background.
The contributions of each model are weighted by $w_\text{X}, w_\text{Y}$, and $w_\text{Z}$ for models X, Y and Z, respectively. 
The signals and the standard model background are mixed at the ratio of $s_0$ to $1$\,{fb} and they are appropriately normalized, 
so that the weights satisfy $\sum w_\text{P} = 1$ with $w_\text{X,Y,Z} \ge 0$.
Thus, the $i$-th bin of the model is written as
\begin{eqnarray}
\label{eq:model}
N_{i,\text{model}}(w_\text{P}) \equiv \sum_{\text{P}} w_\text{P} N_{i\text{P}},
\end{eqnarray}
where P takes two of X, Y or Z.
Hereafter, we denote the number of events in the $i$-th bin of the MC samples, the model P(=one of X,Y or Z), and the fitting model by $N_{i,\text{exp}}$, $N_{i\text{P}}$ and $N_{i,\text{model}}$, respectively, where $i$ takes A, B, $\cdots$ F.

For given MC data sets of the true model, $N_{i,\text{exp}}$, the likelihood function $\lambda(w_\text{P})$ is
\begin{eqnarray}
\label{eq:likelihood}
\ln\lambda(w_\text{P}) = -\sum_{i=\text{A}\cdots\text{F}}\left[ N_{i,\text{model}}(w_\text{P}) - N_{i,\text{exp}} + N_{i,\text{exp}} \ln\frac{N_{i,\text{exp}}}{N_{i,\text{model}}(w_\text{P})}\right],
\end{eqnarray}
if we assume that the distributions follow the Poisson distribution. Detailed discussion on the likelihood function can be found in \cite{Agashe:2014kda}.
We numerically find the zeros of ${\partial\ln\lambda}/{\partial w_\text{P}}$ to obtain the best fit value of $w_\text{P}$. 
In this procedure, $w_\text{X,Y,Z} \ge 0$, is not required for simplicity.

%%%%%%%%%%%%%%%%%%%%%%%%%%%%%%%%%%%%%%%%%%
\begin{figure}[tbp]
        \centering
\includegraphics[width=8cm]{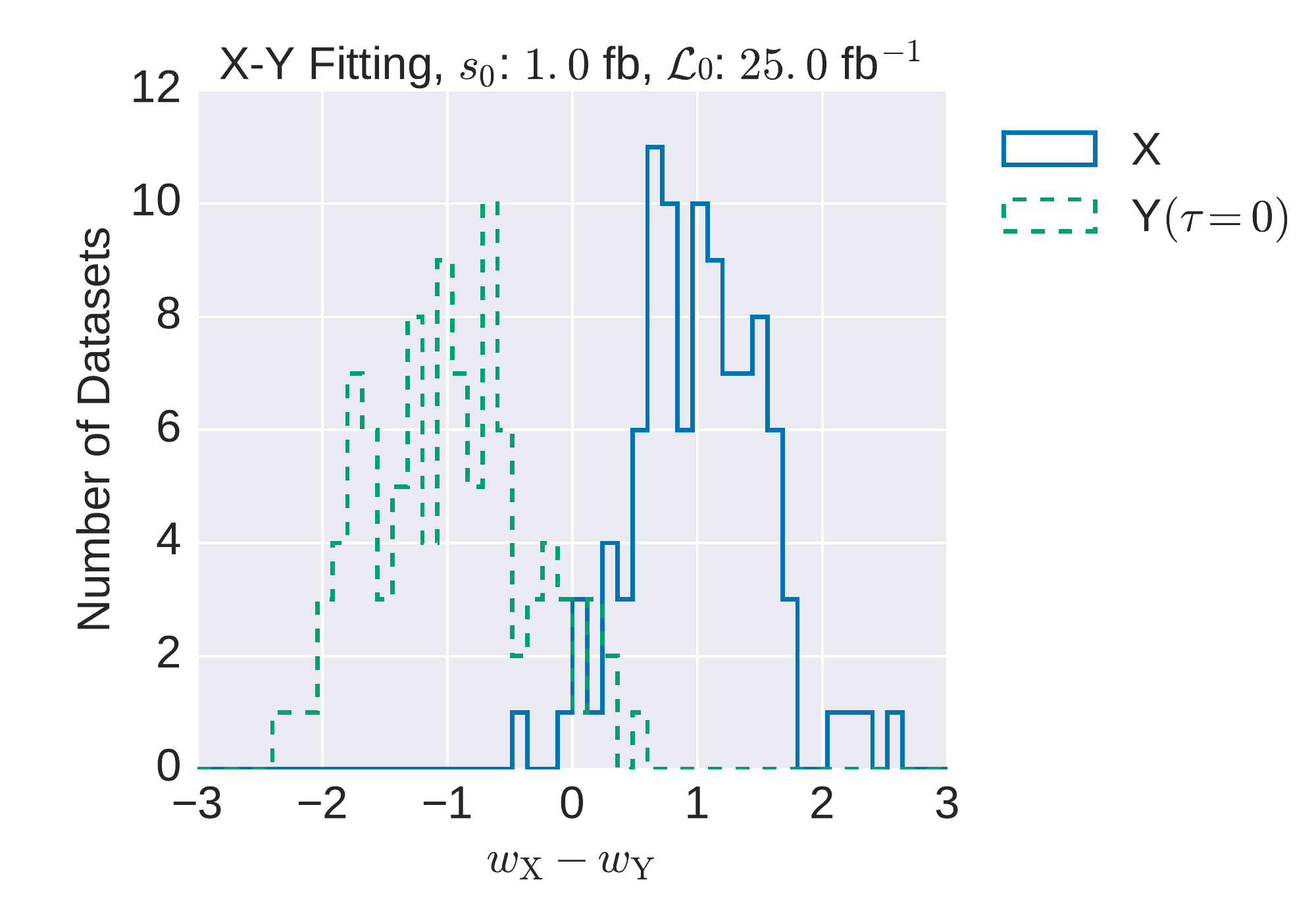}
\includegraphics[width=8cm]{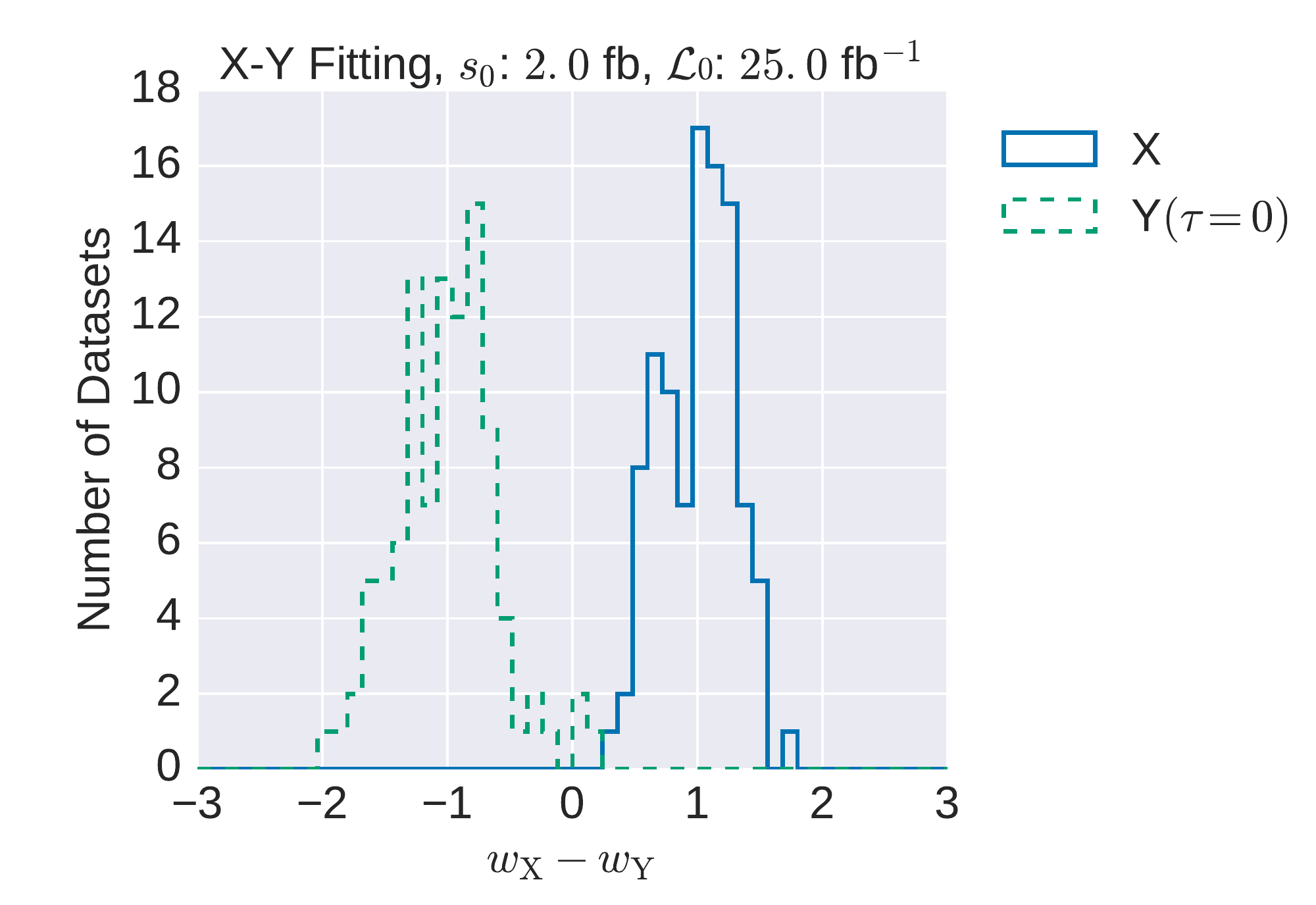}
\includegraphics[width=8cm]{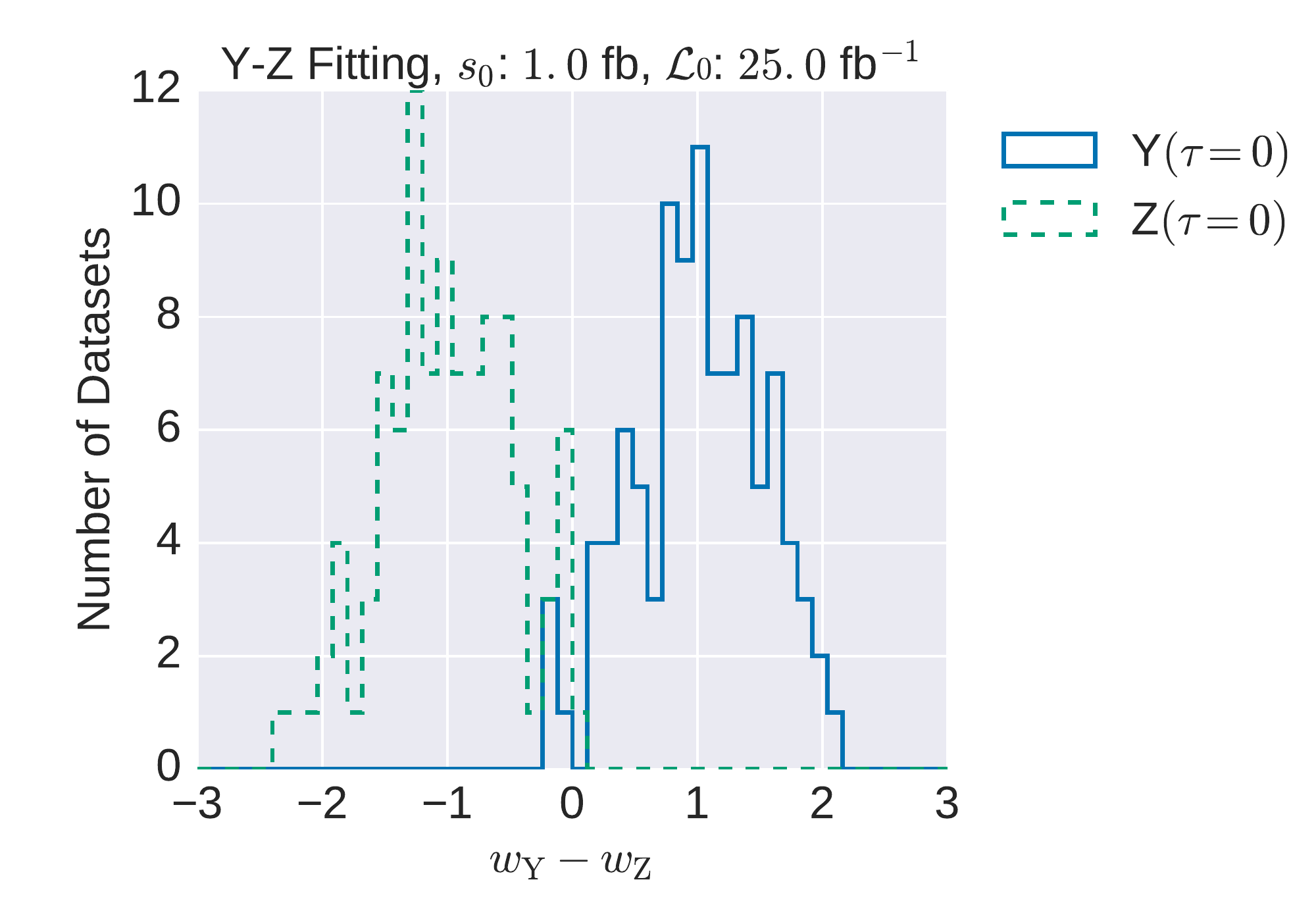}
\includegraphics[width=8cm]{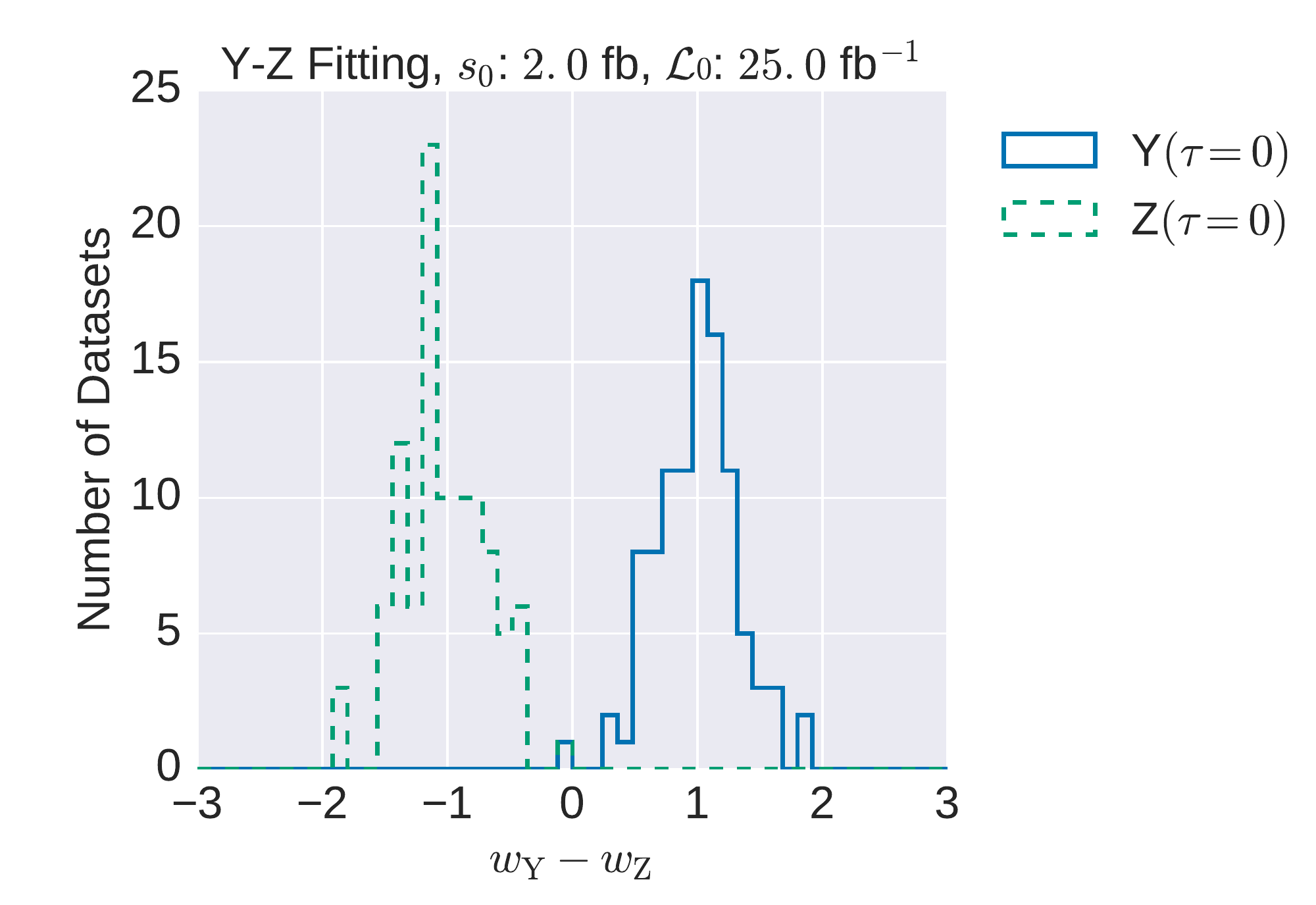}
\includegraphics[width=8cm]{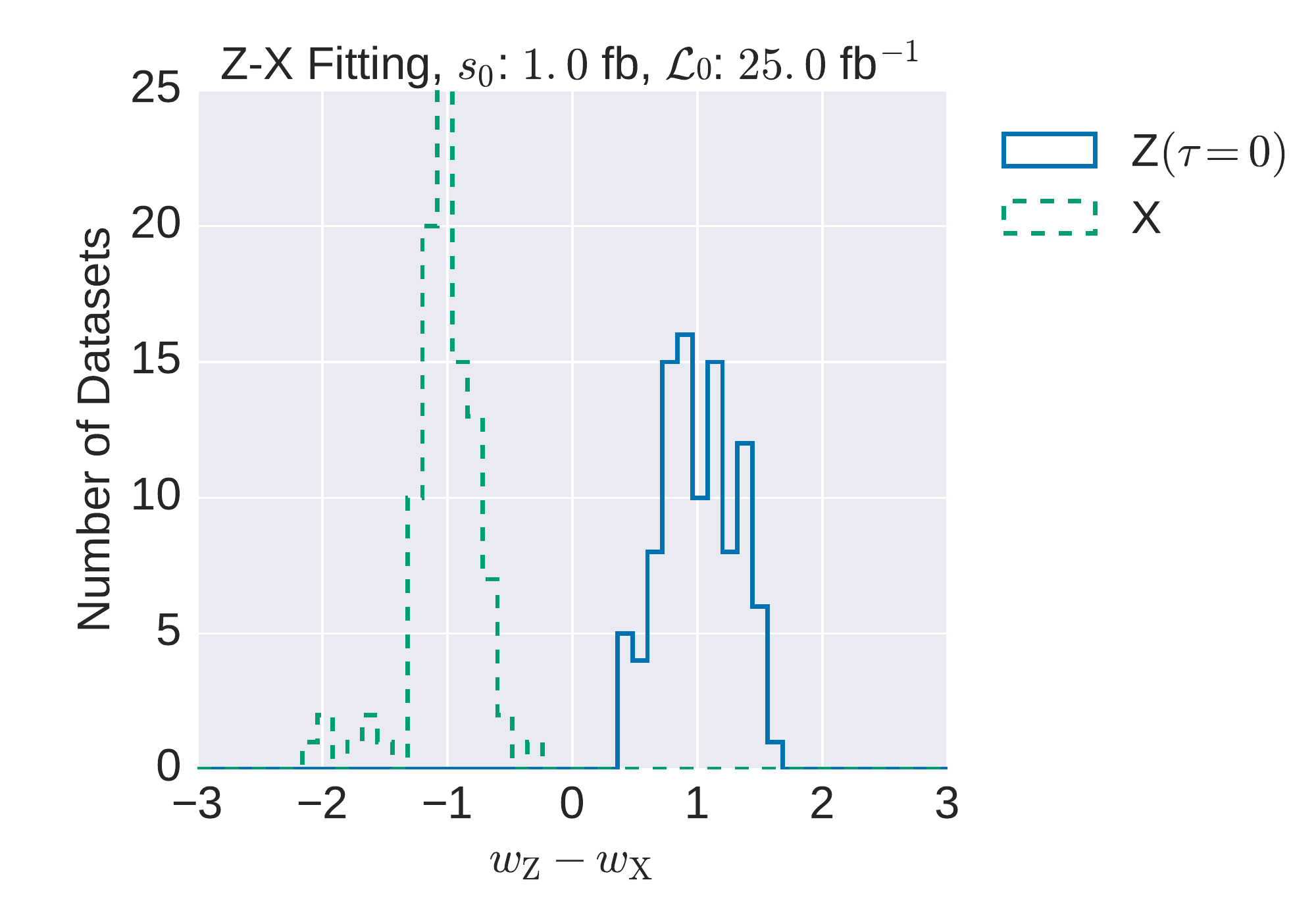}
\includegraphics[width=8cm]{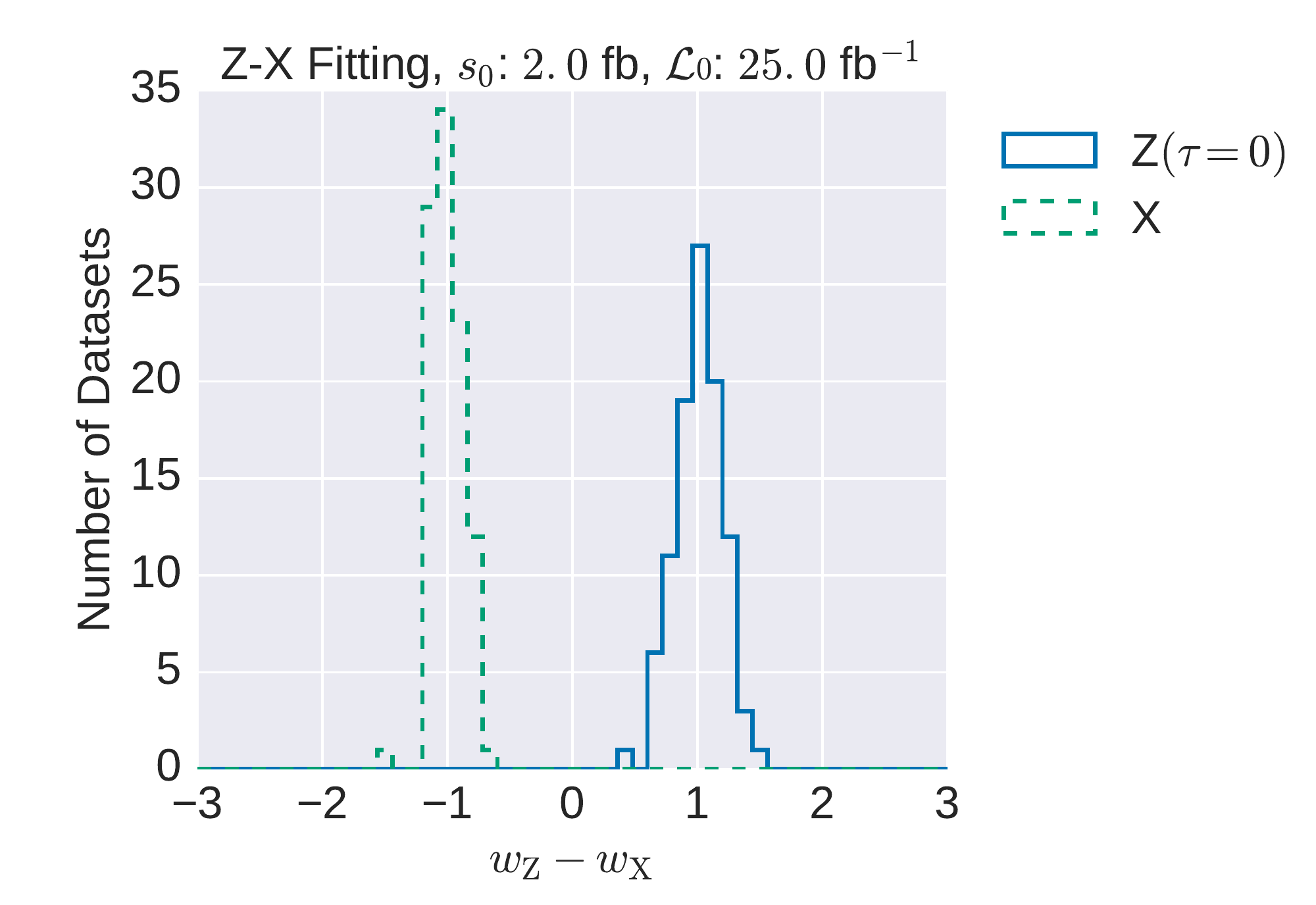}
        \caption{\sl\small The histograms of the fitting parameter $w_\text{X$_1$}-w_\text{X$_2$}$ for each model for a given $s_0$\,fb
        and for ${\cal L}_0 = 25$\,fb$^{-1}$. 
        In each panel, the labels of the histograms on the rights hand side denote the true models.
              }
        \label{fig:chisq25}
\end{figure}
%%%%%%%%%%%%%%%%%%%%%%%%%%%%%%%%%%%%%%%%%%

It should be noted that only $w_\text{P}$ corresponding to the true model is expected to be close to one while the other weights should be small.
To estimate the sensitivity of the experiments, we repeat the pseudo experiments corresponding to the cross section $s_0 = 1$\,fb or $2$\,fb 
for the total luminosity $\mathcal L_0 =25\,\text{fb}^{-1}$. 
Namely, we generate $\mathcal L_0 \times (1\,\text{fb} + s_0) = 50$ and $75$ events as one set of experimental data. 
We also do the same analysis with $s_0 = 0.1\,\text{fb}$ and $\mathcal L_0 = 300\,\text{fb}^{-1}$ in order to cover the low cross section possibility.
In order to roughly estimate the $1\sigma$ region of the distribution of $w_{\rm X, Y, Z}$, we calculate the distribution of the minimal value of $w_\text{P}$ for $100$ 
sets of the MC samples for each model. 
%and thus Eq.\,(\ref{eq:model}) and (\ref{eq:likelihood}) are appropriately modified. 
When models X$_1$ and X$_2$ are taken, the only independent parameter is $w_\text{X$_1$} - w_\text{X$_2$}$. 

\begin{figure}[tbp]
        \centering
        \includegraphics[width=8cm]{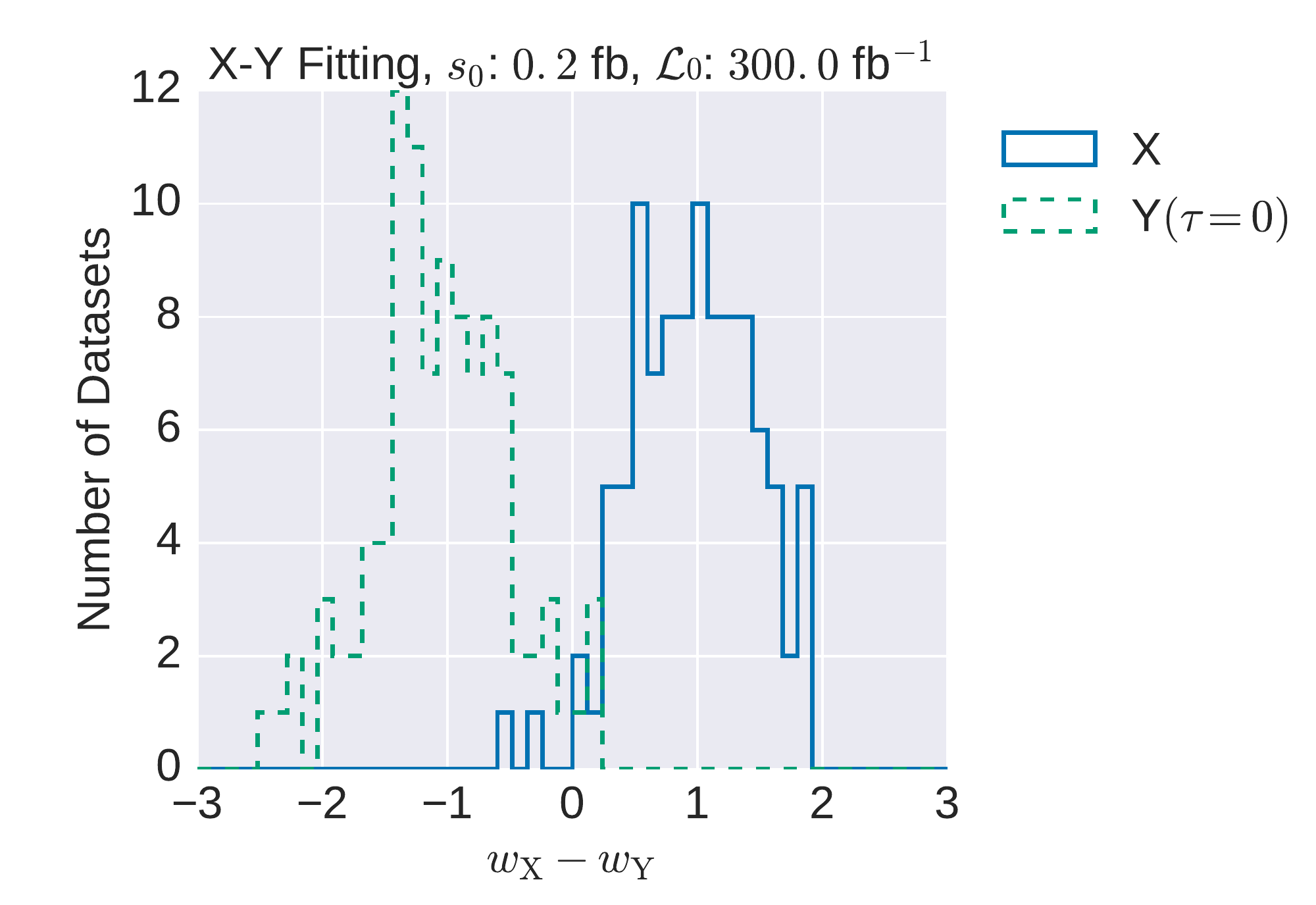}
        \includegraphics[width=8cm]{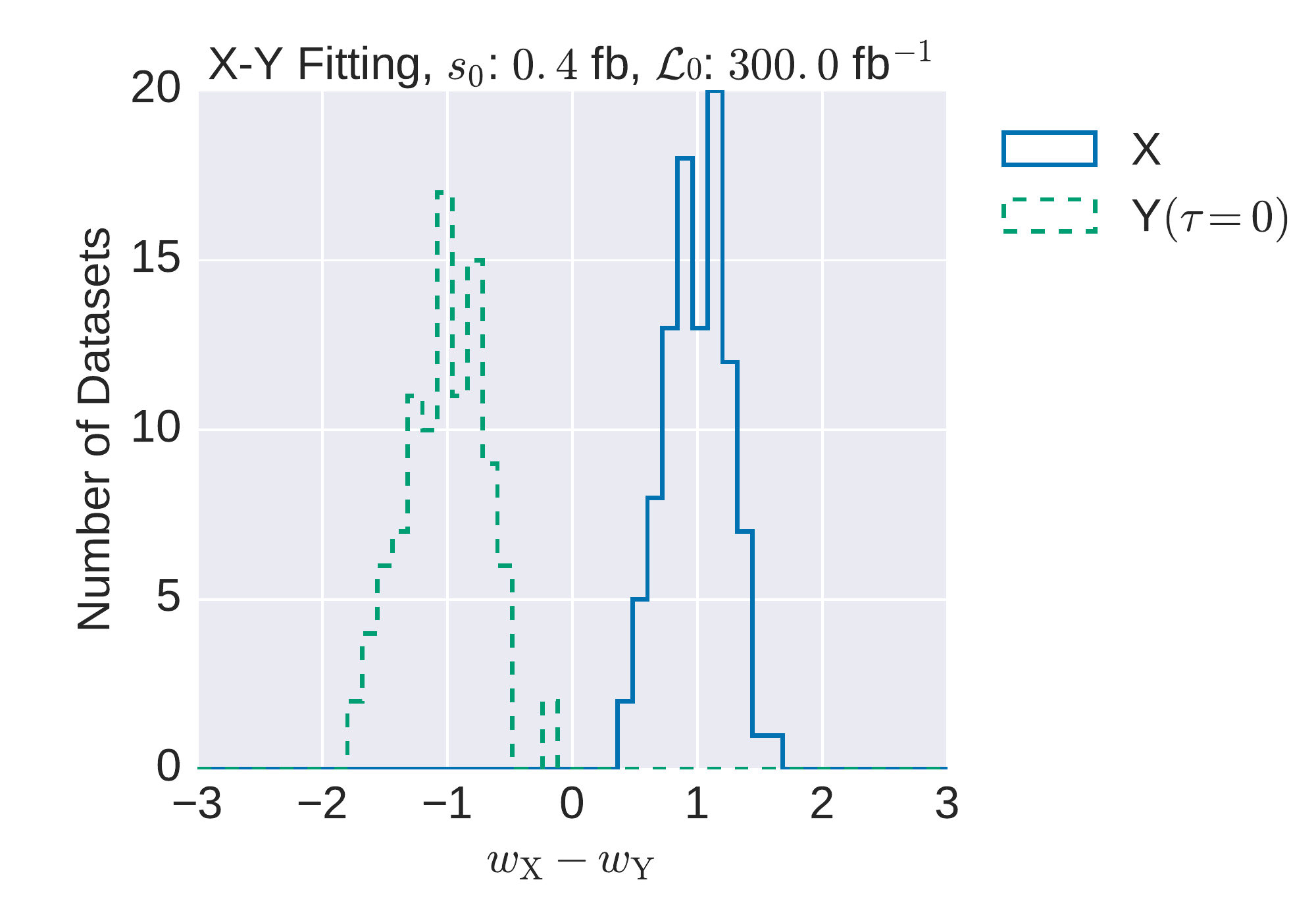}
        \includegraphics[width=8cm]{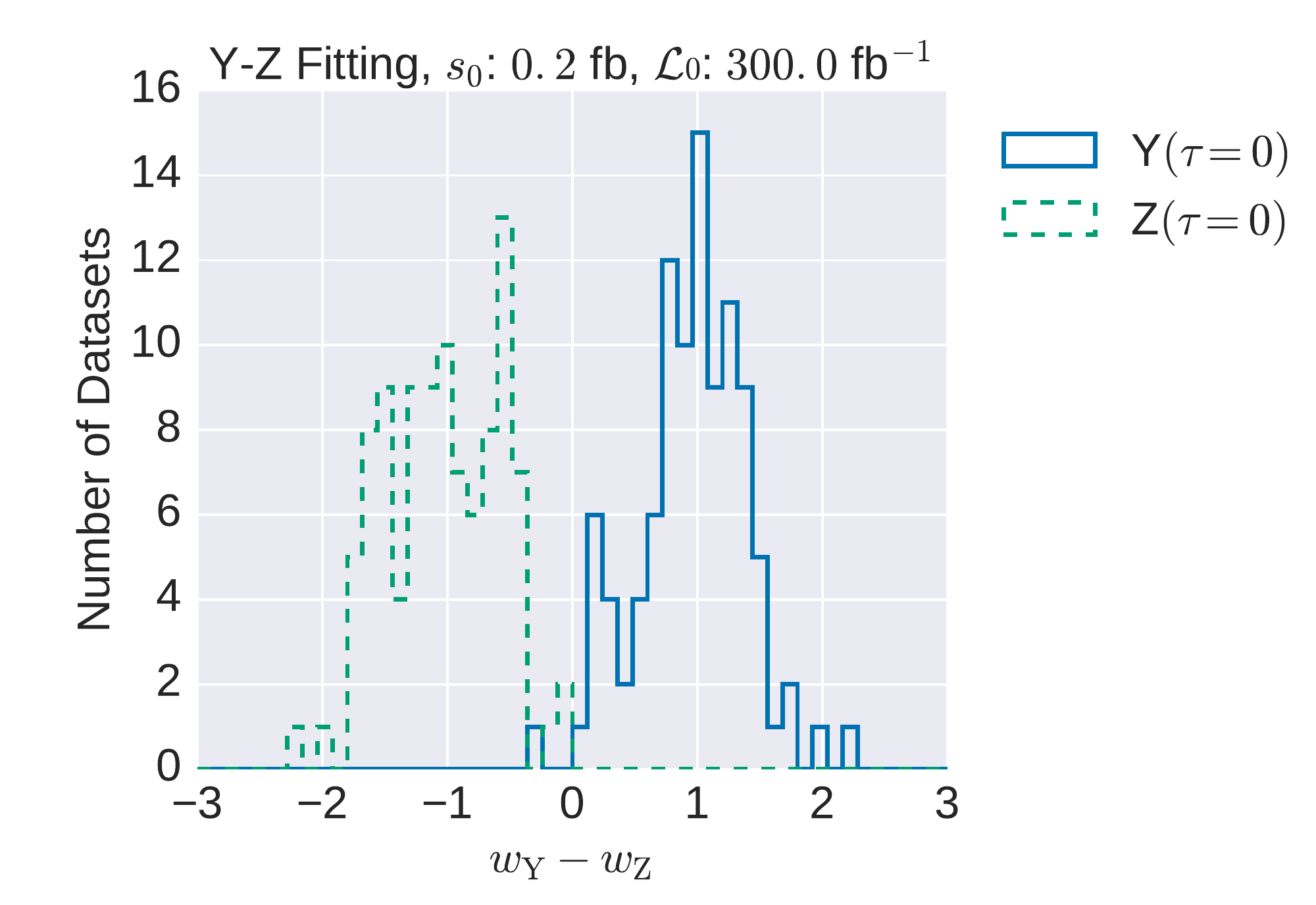}
        \includegraphics[width=8cm]{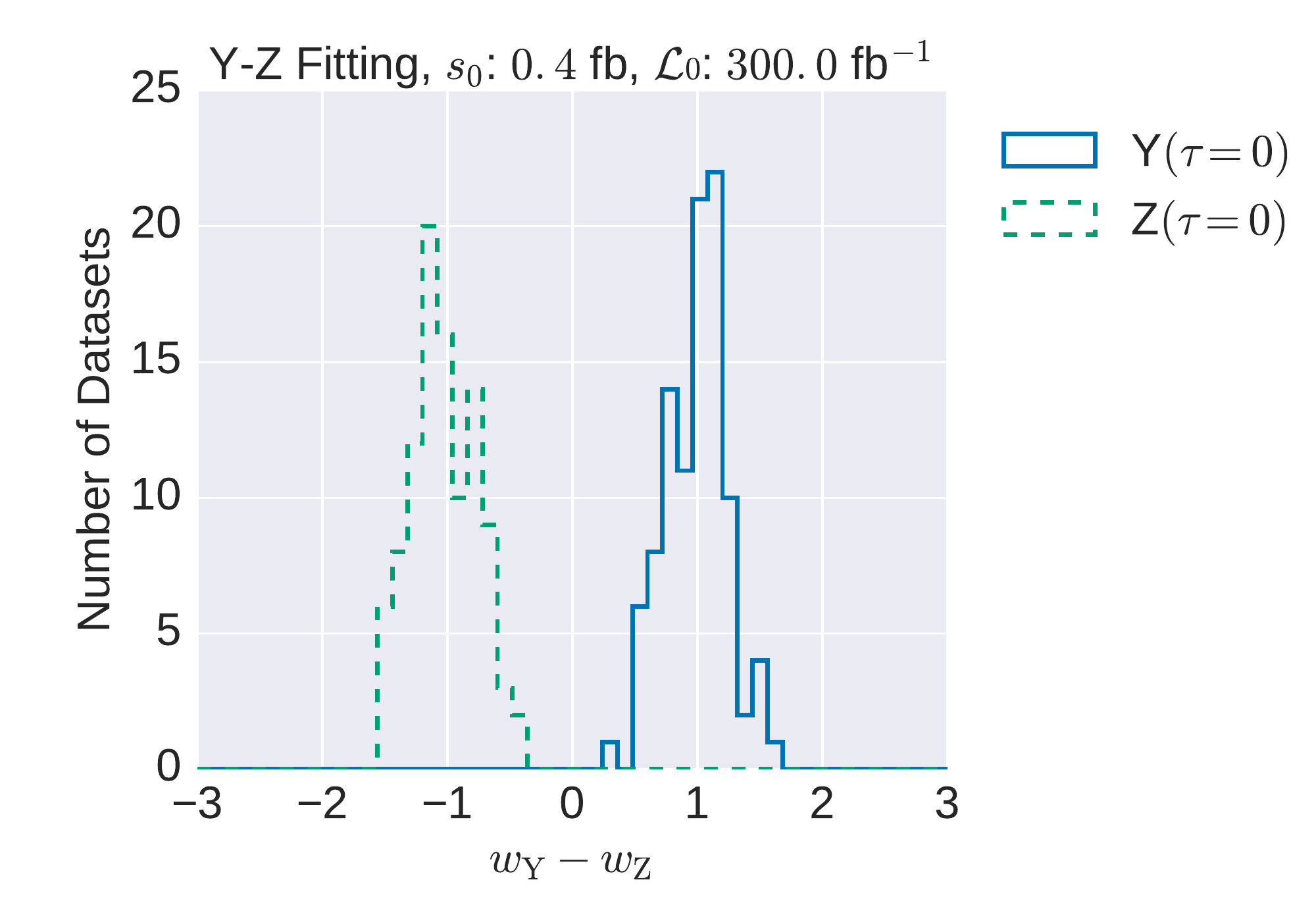}
        \includegraphics[width=8cm]{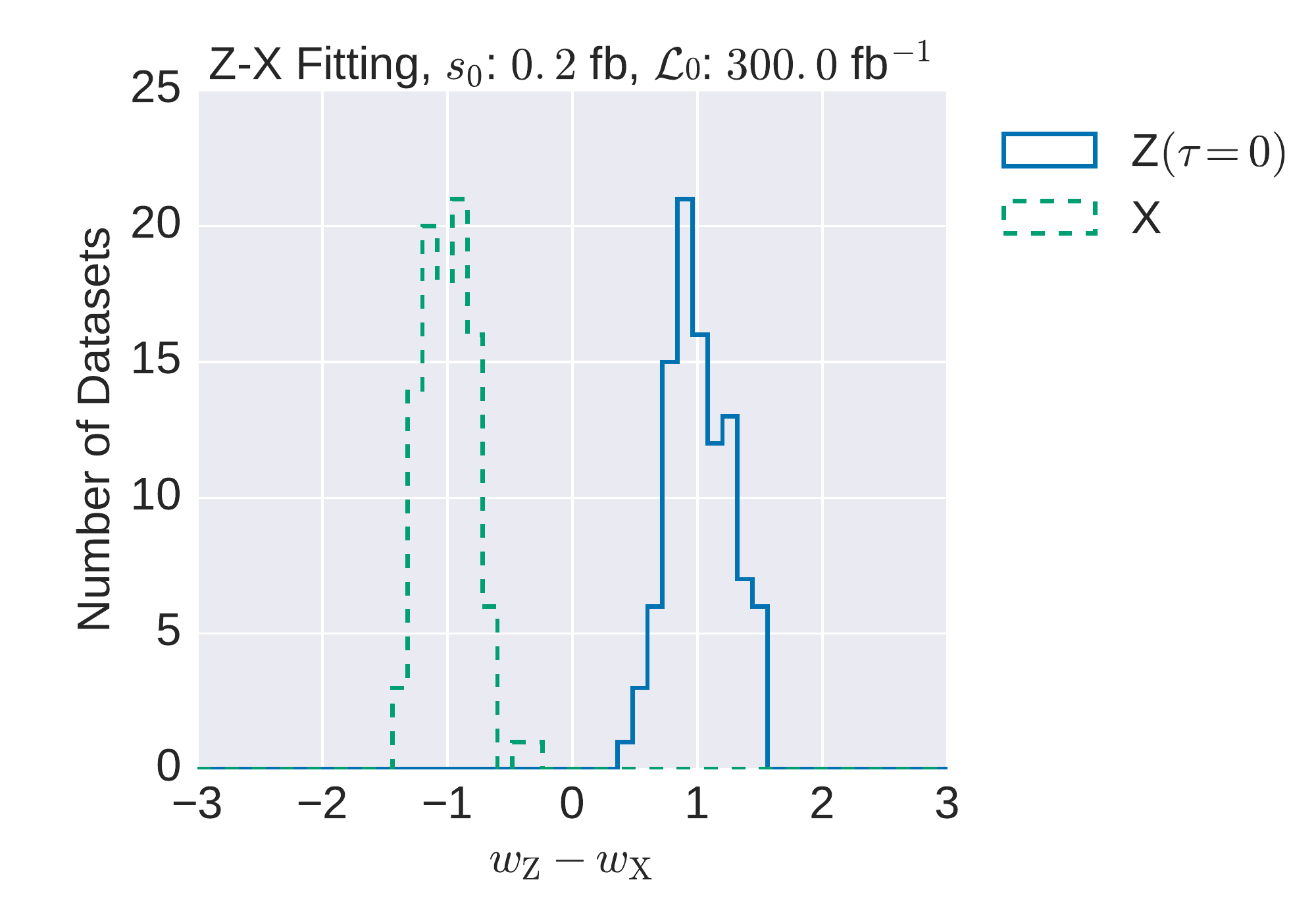}
        \includegraphics[width=8cm]{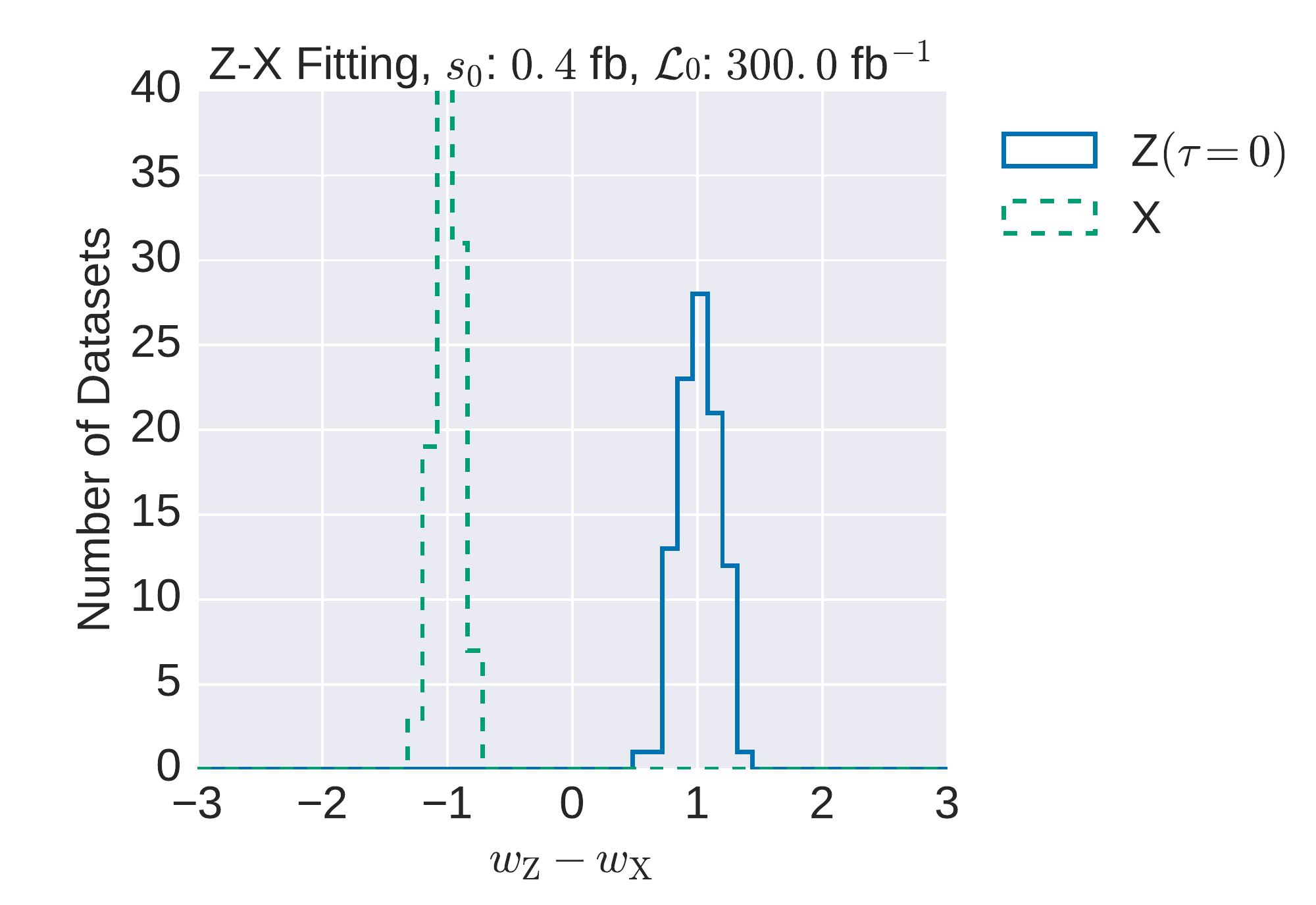}
        \caption{\sl\small The histograms of the fitting parameter $w_\text{X$_1$}-w_\text{X$_2$}$ for each model for a given $s_0$\,fb
        and for ${\cal L}_0 = 300$\,fb$^{-1}$. 
        In each panel, the labels of the histograms on the rights hand side denote the true models.}
        \label{fig:chisq300}
\end{figure}

We show the distribution of the weight difference in Fig.\,\ref{fig:chisq25} and Fig.\,\ref{fig:chisq300}.%
\footnote{As for the signal cross section, $s_0$\,fb, we assume that is is well determined by the 
diphoton signal identified by the ECAL only. } 
Due to numerical instability, some data have been dropped and the total number of events are less than $100$.
In the first row in Fig.\,\ref{fig:chisq25}, we plot the distributions for model X and Y fitted to a mixed model X--Y. There, the distributions for model X and Y are shown as solid and dotted lines, respectively. Similarly, in the second row, model Y and Z are fitted to a mixed model Y--Z, and in the third row, model Z and X are fitted to a mixed model Z--X.
In high statistics limit, the peaks are at $+1$ and $-1$. Since we have allowed negative weights, unphysical small peaks appear where $w_\text{X$_1$} - w_\text{X$_2$} \ne 1$. The distributions are clearly separated for $s_0 = 2\,\text{fb}$.
In the first and second rows, $w_\text{X, Z} - w_\text{Y}$ are less discriminative for model Y.
On the other hand, both weights are well separated for model Z and X in the third row.
In the Fig.\,\ref{fig:chisq300}, we show the distribution with lower cross sections, $0.2$ and $0.4\,\text{fb}$ and higher luminosity, $300\,\text{fb}^{-1}$.

%%%%%%%%%%%%%%%%%%%%%%%%%%%%%%%%%%%%%%%%%%
%%%%%%%%%%%%%%%%%%%%%%%%%%%%%%%%%%%%%%%%%%

We next discuss the significance of our analysis quantitatively. We perform the above analysis with $s_0 = 1\,\text{fb}$ and $\mathcal L_0 = 100\,\text{fb}^{-1}$. Then, the obtained weight distributions are fitted with the Gauss distribution to estimate the mean values and the standard deviations of the weights. The results are shown in Tab.\,\ref{tab:gaus}. In this case, we can distinguish two models with $\sim5\,\sigma$ level.

\begin{table}[tbp]
        \centering
        \caption{\sl\small The Gaussian fitting of the distributions for given models with $s_0=1\,\text{fb}$ and $\mathcal L_0 = 100\,\text{fb}^{-1}$. In the (X$_1$, X$_2$) cell, the mean values and the standard deviations of the weight $w_\text{X$_1$}$ and $w_\text{X$_2$}$ is shown, where the Monte Carlo data from model X$_1$ is fitted with a mixed model of model X$_1$ and X$_2$.
        }
\begin{longtable}[]{@{}c|c|c|c@{}}
\toprule
& X & Y & Z\tabularnewline
\midrule\hline
\endhead
X & & \shortstack{(1.01, 0.10)\\ (-0.01, 0.10)} &
\shortstack{(1.00, 0.04)\\ (0.00, 0.04)}\tabularnewline\hline
Y & \shortstack{(1.01, 0.18)\\ (-0.01, 0.18)} & &
\shortstack{(1.01, 0.12)\\ (-0.01, 0.12)}\tabularnewline\hline
Z & \shortstack{(0.99, 0.07)\\ (0.01, 0.07)} &
\shortstack{(0.99, 0.12)\\ (0.01, 0.12)} &\tabularnewline
\bottomrule
\end{longtable}
        \label{tab:gaus}
\end{table}

Before closing this section, let us comment on the possibility to distinguish the photon-jets from the single photons
solely by the conversion probability as discussed in \cite{Ellwanger:2016qax,Dasgupta:2016wxw}.
As we see from the column F in Tab.\,\ref{tab:binDist}, the differences of the non-conversion rates are not very 
significant, and hence, the discrimination requires rather large integrated luminosity for $s_0 \lesssim 1\,\text{fb}$.
One can see from Fig.\,\ref{fig:chisq25} and Fig.\,\ref{fig:chisq300} that the $p_T$ sum of the first $e^+e^-$ pair 
has more sensitivity to distinguish photon-jets from photons even for $s_0 = {\cal O}(0.1)$\,fb
for ${\cal L}_0 = 300$\,fb$^{-1}$ than the analysis where only the conversion probability is used.

\section{Conversion Length Analysis}
\label{sec:lifetime}
\begin{figure}[tbp]
	\centering
\begin{minipage}{.46\linewidth}
  \includegraphics[width=\linewidth]{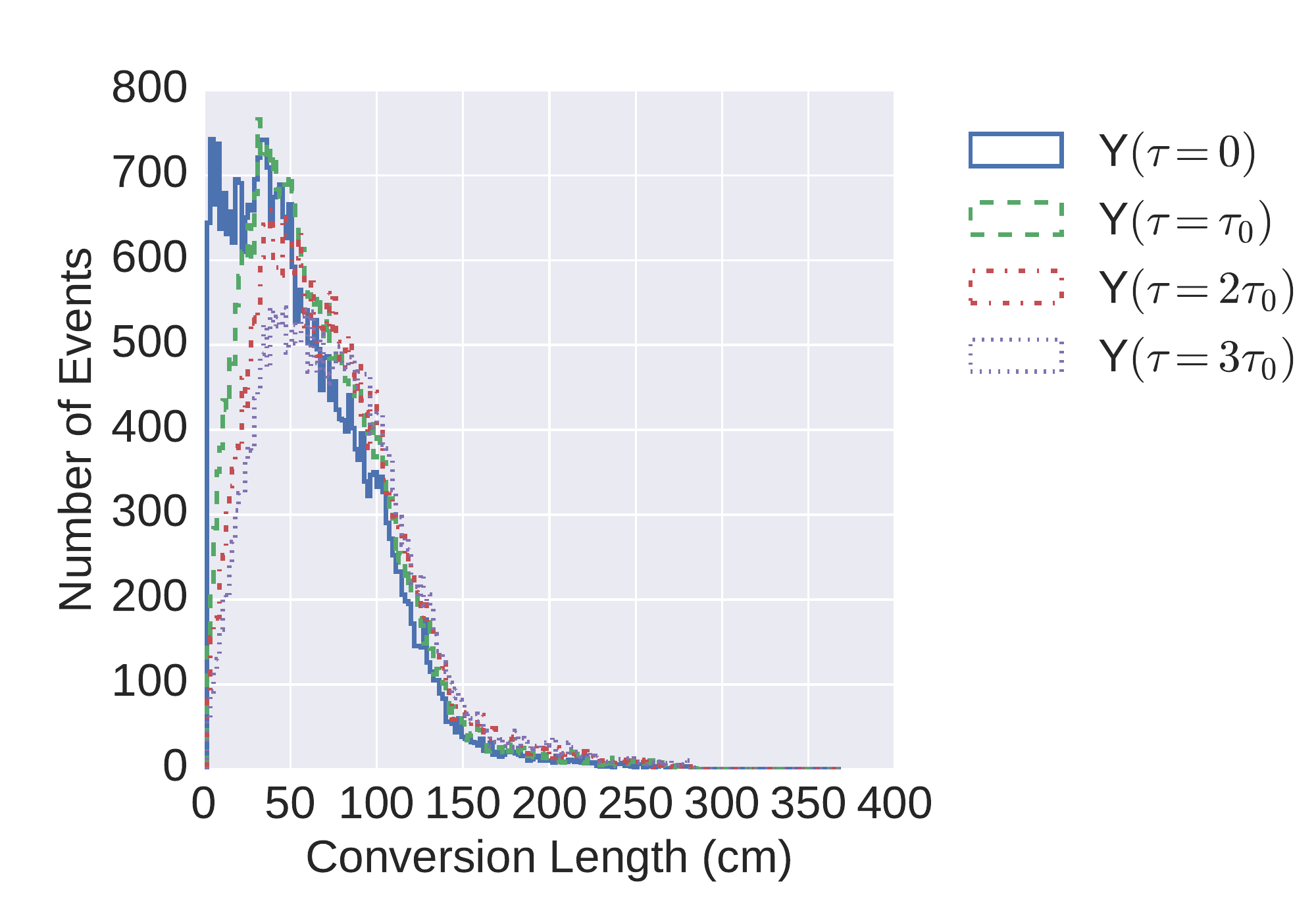}
 \end{minipage}
 \hspace{1cm}
 \begin{minipage}{.46\linewidth}
  \includegraphics[width=\linewidth]{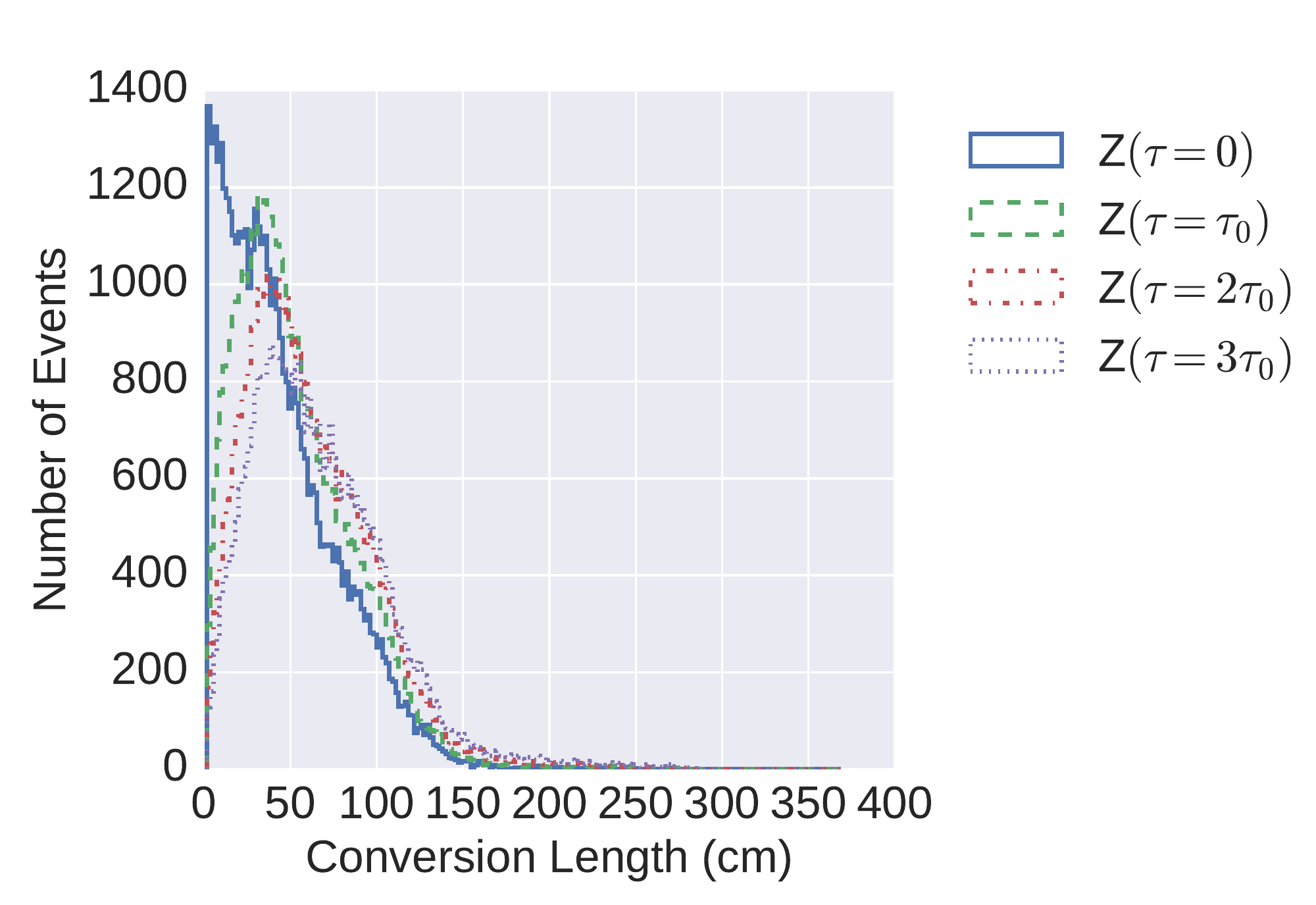}
 \end{minipage}
        \caption{\sl \small The conversion length for model Y (left) and for Z (right).}
        \label{fig:lConvPi0Interm}
\end{figure}

In this section, we focus on the conversion length of the first converted photon, $L_\text{conv}$, to estimate the lifetime of the intermediate particles.
Assuming the lifetime to be $\tau_0, 2\tau_0, 3\tau_0$ where $\tau_0$ is defined by $\beta c \tau_0 \gamma = 10\,\text{cm}$ with $\gamma = {375\,\text{GeV}}/{0.4\,\text{GeV}}$, 
we show the conversion lengths of the first converted photons for model Y and Z obtained by or Monte Carlo simulation in Fig.\,\ref{fig:lConvPi0Interm}.
Here, we used the same set of events as the previous section.
In Fig.\,\ref{fig:lConvPi0Interm}, the distributions of model Y and Z are different. The conversion point is therefore also sensitive to distinguish the number of photons but we do not use it in our fit.
We again classify the events into following 6 categories;

%%%%%%%%%%%%%%%%%%%%%%%%%%%%%%%%%%%%%%%%%%%%%%%%%%%%%%%%
%%%%%%%%%%%%%%%%%%%%%%%%%%%%%%%%%%%%%%%%%%%%%%%%%%%%%%%%

\begin{description}
\item[P] Events with $L_\text{conv} \le 10\,\text{cm}$
\item[Q] Events with $10\,\text{cm} < L_\text{conv} \le 20\,\text{cm}$
\item[R] Events with $20\,\text{cm} < L_\text{conv} \le 30\,\text{cm}$
\item[S] Events with $30\,\text{cm} < L_\text{conv}$
\item[T] Events with no photon being converted by the end of the tracker. 
\item[U] Events with the intermediate particles decaying outside of the inner trackers.
\end{description}
The fractions of events falling into these categories are given in Tab.\,\ref{tab:lConvDist}, which are obtained in a similar way as those in Tab.\,\ref{tab:binDist}.
The category U could lead to striking signatures if the intermediate particle decays inside the ECAL. Such a event would appear as an isolated high-energy ECAL activity at the middle of the electromagnetic calorimeter.
For $\tau \lesssim 3\tau_0$, however, the expected fraction is highly suppressed. 
% {\color{red} In the subsequent analysis, we drop the events categorized into U, since the events are not identified 
% with the diphoton signal by the ECAL. ?}

%%%%%%%%%%%%%%%%%%%%%%%%%%%%%%%%%%%%%%%%%%%%%%%%%%%%%%%%
\begin{table}[tbp]
        \centering
        \caption{\sl \small The distributions of the classification above. ``SM BG'' corresponds to the standard model background.
        }
\begin{longtable}[]{@{}c|cccccc@{}}
\toprule
& P & Q & R & S & T & U\tabularnewline
\midrule\hline
\endhead
X & 0.037 & 0.036 & 0.036 & 0.345 & 0.547 & 0.000\tabularnewline\hline
Y\((\tau = 0)\) & 0.070 & 0.066 & 0.070 & 0.464 & 0.329 &
0.000\tabularnewline
Y\((\tau = \tau_0)\) & 0.022 & 0.049 & 0.062 & 0.511 & 0.356 &
0.000\tabularnewline
Y\((\tau = 2\tau_0)\) & 0.013 & 0.032 & 0.045 & 0.513 & 0.393 &
0.004\tabularnewline
Y\((\tau = 3\tau_0)\) & 0.010 & 0.024 & 0.038 & 0.492 & 0.415 &
0.021\tabularnewline\hline
Z\((\tau = 0)\) & 0.137 & 0.118 & 0.110 & 0.496 & 0.140 &
0.000\tabularnewline
Z\((\tau = \tau_0)\) & 0.046 & 0.088 & 0.106 & 0.596 & 0.165 &
0.000\tabularnewline
Z\((\tau = 2\tau_0)\) & 0.026 & 0.058 & 0.081 & 0.632 & 0.199 &
0.005\tabularnewline
Z\((\tau = 3\tau_0)\) & 0.018 & 0.044 & 0.066 & 0.622 & 0.229 &
0.022\tabularnewline\hline
SM BG & 0.040 & 0.038 & 0.039 & 0.385 & 0.500 & 0.000\tabularnewline
\bottomrule
\end{longtable}
        \label{tab:lConvDist}
\end{table}
%%%%%%%%%%%%%%%%%%%%%%%%%%%%%%%%%%%%%%%%%%%%%%%%%%%%%%%%

\begin{figure}[tbp]
        \centering
        \includegraphics[width=8cm]{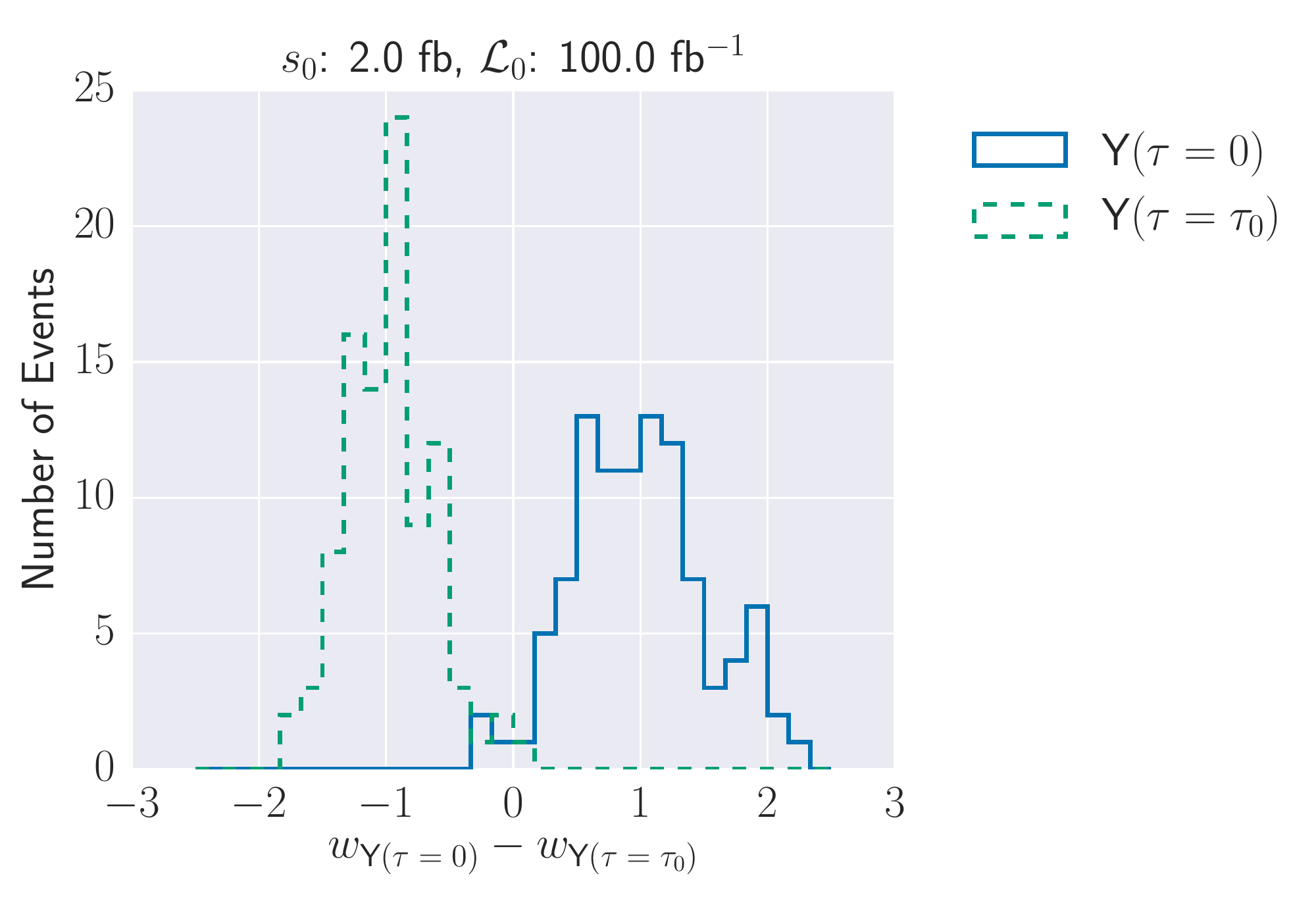}
        \includegraphics[width=8cm]{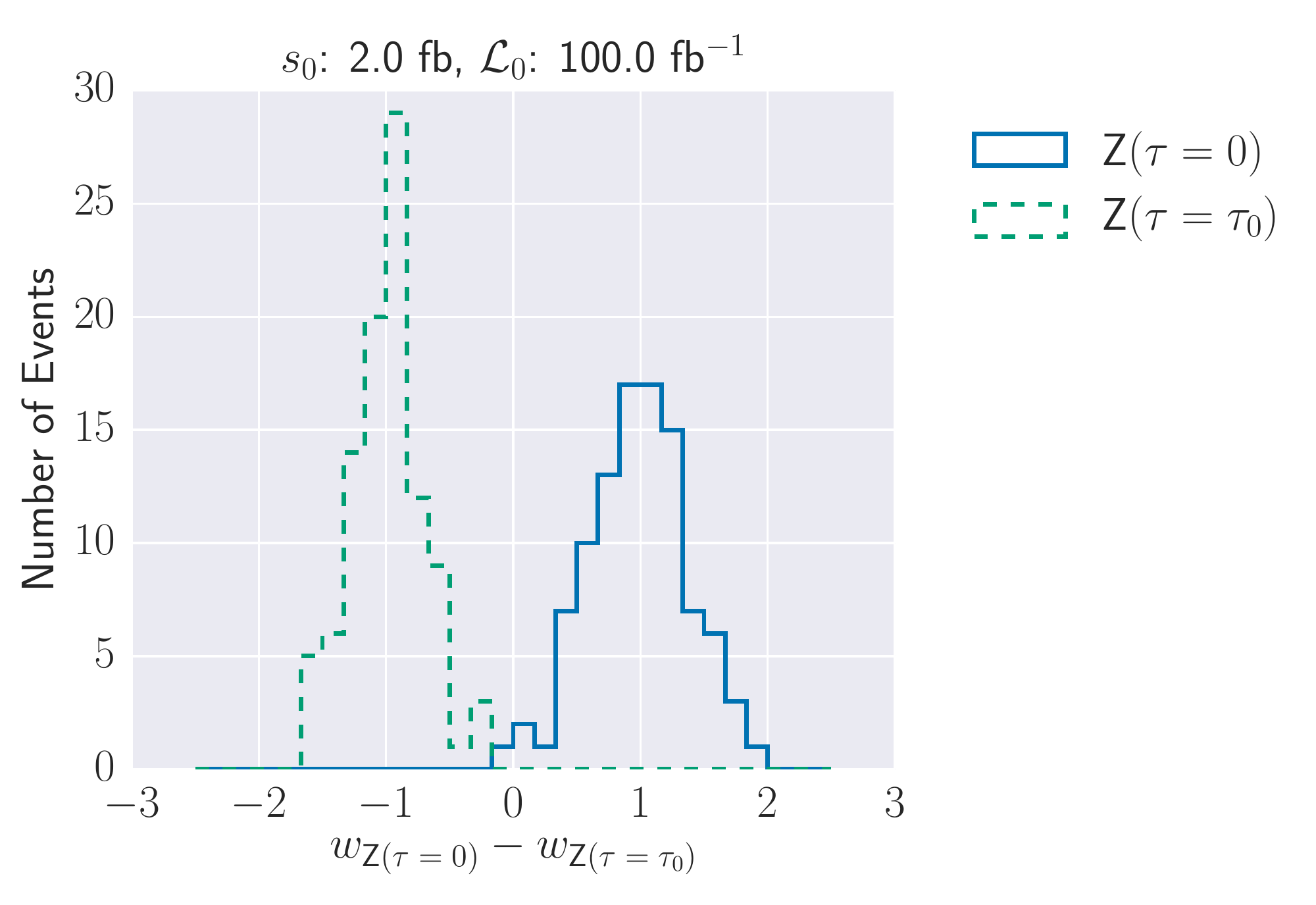}
        \includegraphics[width=8cm]{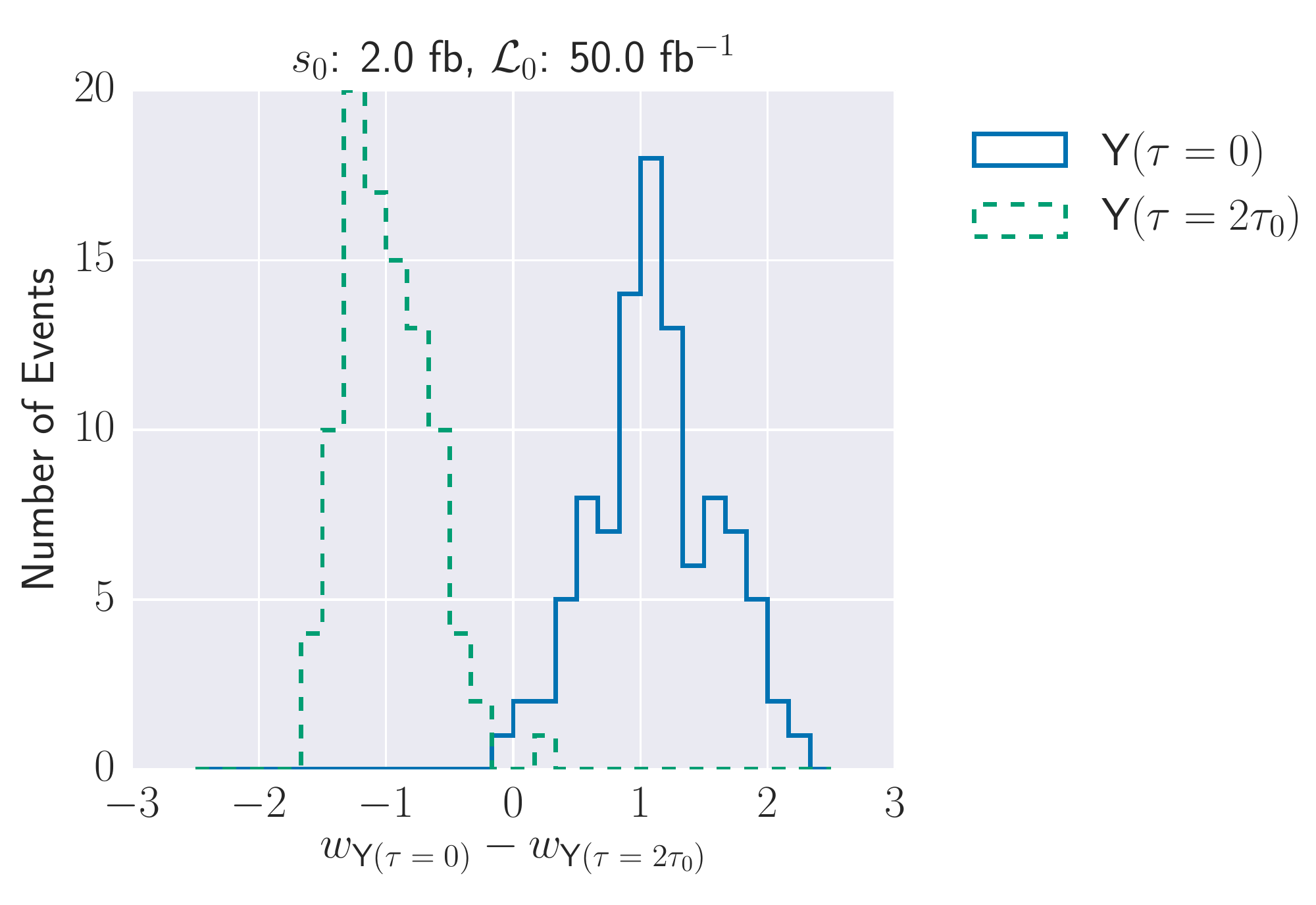}
        \includegraphics[width=8cm]{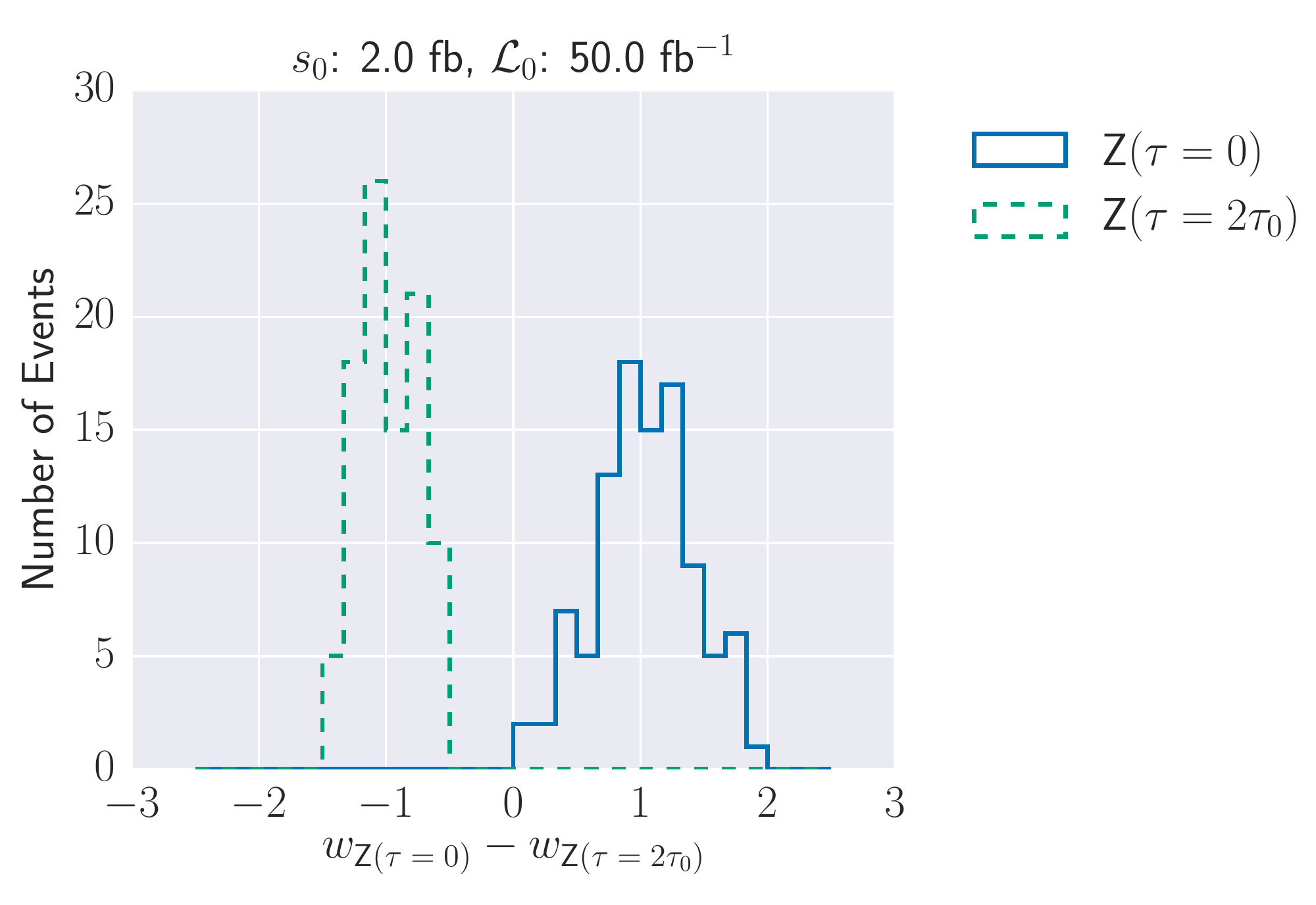}
        \includegraphics[width=8cm]{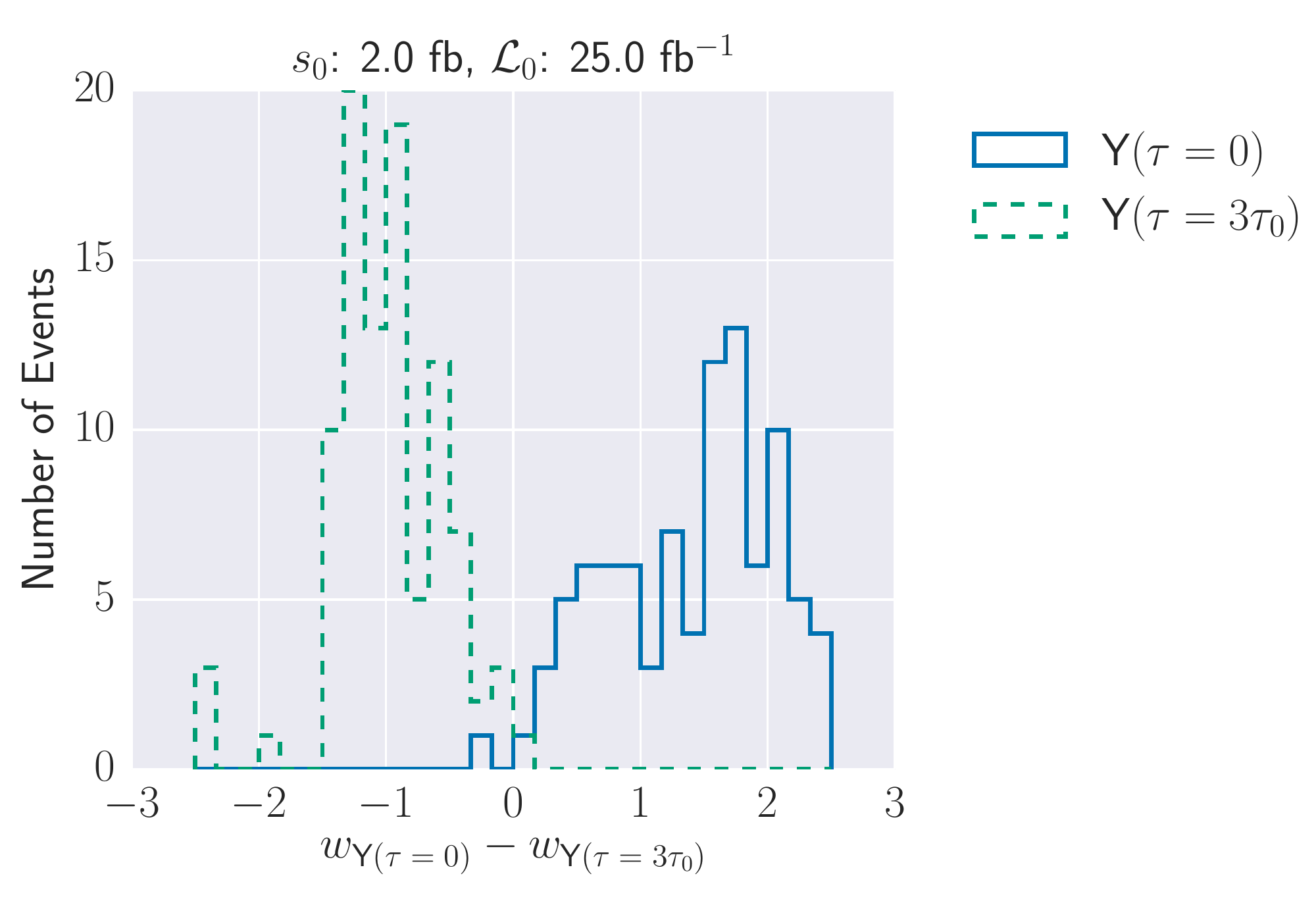}
        \includegraphics[width=8cm]{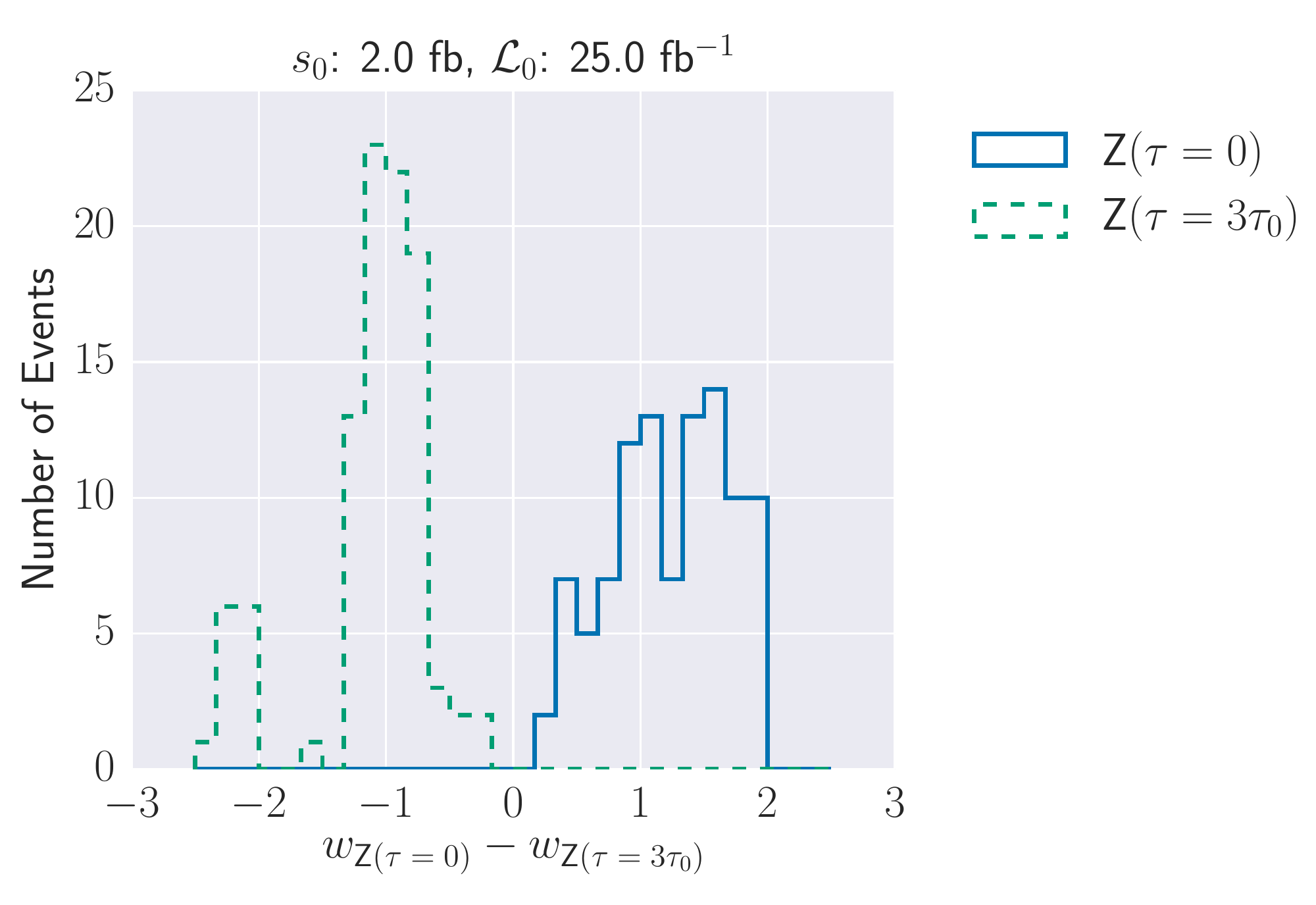}
        \caption{\sl \small
        The comparisons between $\tau=0$ and non-zero $\tau$ based on the conversion length in model Y and Z. 
       The MC data sets, the cross sections and the luminosity are noted as titles.}
              \label{fig:lConv}
\end{figure}

For given MC events, the similar likelihood analysis as the previous section is performed. 
Here we compare the Monte Carlo data with mixed model with different lifetime while the decay model is fixed to either Y or Z. 
The results of the likelihood analyses for the mixed model of $\tau = 0$ and $\tau = \tau_0$, $\tau = 0$ and $\tau = 2\tau_0$ 
and $\tau  = 0$ and $\tau =3\tau_0$ 
for $s_0=2\,\text{fb}$ and $\mathcal L_0=25\,\text{fb}^{-1}$ are shown in Fig.\,\ref{fig:lConv}. 
As is the same as in Fig.\,\ref{fig:chisq25} and \ref{fig:chisq300}, the only independent parameter is the difference of the weights, $w_{\text{P}(\tau=\tau_1)} - w_{\text{P}(\tau=\tau_2)}$. Here, $w_{\text{P}(\tau=\tau_i)}$ denotes the weight of  a model $P$ with lifetime $\tau_i$.
We choose the luminosity $\mathcal L_0$ depending on the lifetime such that both distributions are separated enough.　It is shown that $100, 50$ and $25\,\text{fb}^{-1}$ luminosity is necessary to distinguish $\tau=\tau_0, 2\tau_0$ and $3\tau_0$, respectively.

\label{sec:lifetime}

\section{Conclusions}
In this paper, we discussed how the photon-jet signal can be distinguished from the photon 
signal by taking the $750$\,GeV resonance as an example.
The sum of $p_T$ of the  first $e^+e^-$ pair from the photon conversion
provides strong discrimination power.
As a result, we find that it is possible to discriminate the di-photon-jet resonance from
the diphoton resonance at around $2\sigma$ level with the integrated luminosity ${\cal L}_0 = 25$\,fb$^{-1}$
for a signal cross section after the cut of $1$\,fb.
Furthermore, even for a small production cross section of about $0.2$\,fb, it is still possible
to discriminate the di-photon-jet scenario from the diphoton scenario with the integrated luminosity ${\cal L}_0 = 300$\,fb$^{-1}$.

We also discussed the possibility to measure the lifetime of the intermediate particles
of ${\cal O}(10)$\,cm by the distribution of the photon conversion length.
We found that it might be possible
to distinguish models with $\tau = 3\tau_0$ from $\tau = 0$ for $s_0 =2$\,fb with ${\cal L}_0 = 25$\,fb$^{-1}$,
models with $\tau = 2\tau_0$ from $\tau = 0$ for $s_0 =2$\,fb with ${\cal L}_0 = 50$\,fb$^{-1}$ and 
models with $\tau = \tau_0$ from $\tau = 0$ for $s_0 =2$\,fb with ${\cal L}_0 = 100$\,fb$^{-1}$. 

In our estimation, we took into account the detector geometry and the effect of bremsstrahlung in electron reconstruction.
However, it should be noted that we have not considered the electron and photon reconstruction efficiency in our estimation since we only have limited information about the detector and performance.
For more refined analysis, we need more detailed detector simulations, which is out of the scope of our paper.

 % comment wide width
 % comment chiral symmetry breaking

 % So far, not detailed analysis has been done
 % Only good paper is ....
 % In this paper, we try to crack down on the events with fake photons, 
 % where ... and ... and 

 % It should be also noted that the typical decay length of the intermediate particle can be long.
 % As it is discussed in the good paper, 
 % We also address how we can extract the finite decay length of the 

%ALTERNATIVE 
%%%%%%%%%%%%%%%%%%%%%%%%%%%%%%%%%%%%%%%
%%%%%%%%%%% Acknowledgments %%%%%%%%%%%
%%%%%%%%%%%%%%%%%%%%%%%%%%%%%%%%%%%%%%%
\begin{acknowledgments}
The authors thank Takeo Moroi for pointing out errors in our calculation.
This work is supported in part by Grants-in-Aid for Scientific Research from the Ministry of Education, Culture, Sports, Science, and Technology (MEXT), Japan, No. 25105011, No. 15H05889 (M. I.) and No. 23104006 (M. N.); Grant-in-Aid No. 26287039 (M. I. and M. N.), No. 16H03991 (M. N.) and No. 16H06489 (O. J.) from the Japan Society for the Promotion of Science (JSPS); and by the World Premier International Research Center Initiative (WPI), MEXT, Japan (M. I.). 
The work of H.F. is supported in part by a Research Fellowship for Young Scientists from
the Japan Society for the Promotion of Science (JSPS).
\end{acknowledgments}
%%%%%%%%%%%%%%%%%%%%%%%%%%%%%%%%%%
%%%%%%%%%%% References %%%%%%%%%%%
%%%%%%%%%%%%%%%%%%%%%%%%%%%%%%%%%%
% \begin{thebibliography}{99}
% \input{Reference.bib}
% \end{thebibliography}
%\bibliography{../../papers}
\bibliography{papers}

\end{document}